# Parameter-Efficient Continuous-Variable Photonic Quantum Neural Networks for Edge Quantum AI: Demonstration in Oral Cancer Detection

**Akshay Bhagwan Sonawane*, Sophie Choe, and Lakshman Tamil**

Dept. of Electrical and Computer Engineering
University of Texas at Dallas
Richardson, TX 75080

## Abstract

Early detection of oral cancer markedly improves clinical outcomes, yet specialized diagnostic tools remain scarce in low-resource settings. Smartphone-based screening is a scalable alternative but needs lightweight models that run within edge-hardware constraints. Hybrid classical–quantum architectures are emerging candidates for parameter-efficient learning, yet most rely on qubit hardware that needs cryogenic operation, unsuitable for edge deployment. Continuous-variable (CV) photonic quantum computing, which operates at room temperature, offers a complementary route. We investigate a hybrid classical-CV quantum classifier for oral cancer detection from smartphone images. The pipeline combines a MobileNetV1 feature extractor, principal component analysis to 16 dimensions, and a parameterized CV-QNN of displacement, interferometric, and Kerr gates on a photonic backend. We propose a simplified $\Phi \circ D \circ U_1$ CV-QNN architecture that cuts trainable parameters 40-45% relative to the standard CV-QNN layer of Killoran et al. (2019a), and identify dimensionality-reduction and encoding-restriction strategies that mitigate barren plateaus, raising loss-gradient variance by roughly 58 orders of magnitude. Whether the simplified layer beats the full layer is width-dependent: the full layer holds a small but significant edge at two qumodes, whereas the simplified layer is significantly better at four qumodes using 44% fewer parameters. The strongest model, a four-qumode simplified CV-QNN with only 18 parameters, attains the highest validation AUC of all models, exceeds a 55-parameter classical baseline using 67% fewer parameters, and reaches 100% calibrated test accuracy across all seeds. These results support CV photonic quantum machine learning for parameter-efficient, room-temperature medical image classification and motivate progress toward edge quantum AI.

## 1. Introduction

Oral cancer is a major global health concern, with survival rates strongly dependent on early diagnosis (Jubair et al. 2022; Swamikannan et al. 2024). Despite advances in medical imaging and artificial intelligence, timely detection remains challenging, particularly in low-resource and remote settings where access to specialized equipment and trained clinicians is limited (Uthoff et al. 2018; Warin et al. 2021). The widespread availability of smartphones equipped with high-resolution cameras presents an opportunity to democratize oral cancer screening by enabling image-based diagnostics at the point of care (Uthoff et al. 2018; Swamikannan et al. 2024). However, deploying accurate and robust machine learning models on edge devices requires architectures that balance predictive performance, computational efficiency, and energy constraints (Howard et al. 2017; Sze et al. 2017).

Convolutional neural networks (CNNs) have demonstrated strong performance in medical image analysis, including oral lesion classification, but many state-of-the-art models are computationally expensive and unsuitable for edge deployment (Jubair et al. 2022; Swamikannan et al. 2024). Lightweight architectures such as MobileNetV1 address this limitation by employing depthwise separable convolutions, significantly reducing model size and inference cost while maintaining competitive accuracy (Howard et al. 2017). As a result, MobileNetV1 is well suited for processing smartphone-acquired images directly on edge devices, enabling low-latency inference and enhanced data privacy (Sze et al. 2017; Swamikannan et al. 2024). Nonetheless, purely classical models may encounter representational limitations when operating under strict resource constraints or when faced with highly nonlinear feature distributions common in biomedical imagery (Havlíček et al. 2019; Kulkarni et al. 2023).

* akshaybhagwan.sonawane@utdallas.edu

Quantum machine learning has emerged as a promising paradigm for enhancing classical learning models by exploiting the high-dimensional Hilbert space and intrinsic nonlinearity of quantum systems (Schuld and Killoran 2019; Havlíček et al. 2019). Hybrid classical–quantum architectures, in which classical neural networks interface with quantum circuits, are particularly attractive for near-term applications, as they allow quantum components to augment rather than replace existing classical pipelines (Killoran et al. 2019a; Bharti et al. 2022; Kulkarni et al. 2023). Recent studies have shown that quantum neural networks can improve expressivity and generalization for classification tasks, even when operating on a small number of encoded features (Schuld et al. 2020; Kulkarni et al. 2023).

Among quantum computing paradigms, continuous-variable quantum computing is especially suitable for near-term applications (Killoran et al. 2019a). Unlike qubit-based systems that rely on isolating discrete two-level quantum states (Braunstein and Van Loock 2005) and typically require cryogenic environments (Havlíček et al. 2019), continuous-variable systems operate on bosonic modes and leverage the full spectrum of quantum energy levels (Killoran et al. 2019b). Implemented using quantum optical components such as displacement, squeezing, and interferometric gates (Bangar et al. 2022), continuous-variable quantum processors can operate at room temperature (Arrazola et al. 2021). Photonic platforms, such as those developed by Xanadu (Arrazola et al. 2021), provide a practical and scalable foundation for continuous-variable quantum neural networks (Killoran et al. 2019a). Together, room-temperature operation, photonic implementation, and a scalable architecture position continuous-variable quantum computing as a particularly promising candidate for near-term, edge AI, low-power quantum machine learning, motivating the present feasibility study.

In this work, we propose a classical-quantum hybrid oral cancer detection framework that combines MobileNetV1, principal component analysis, and a continuous-variable quantum neural network implemented on a photonic quantum processing unit. MobileNetV1 serves as an efficient feature extractor for smartphone-acquired oral images and is trained using a classical Bayesian neural network, after which its learned parameters are transferred to the hybrid model. Principal component analysis is applied to compress the extracted features into a low-dimensional representation suitable for quantum encoding. The resulting features are encoded into a continuous-variable quantum circuit using displacement and squeezing gates and processed by a trainable quantum neural network optimized via standard backpropagation (Schuld et al. 2015; Killoran et al. 2019a). By integrating edge-efficient classical learning with room-temperature quantum processing, this approach anticipates a future in which quantum parameter amplifiers and photonic quantum processors are co-deployed with classical inference hardware in edge devices.

The contributions of this paper are as follows: (1) the design and empirical evaluation of a hybrid classical–CV-quantum classifier for oral cancer detection from smartphone-acquired images, demonstrating feasibility on a clinically meaningful medical imaging task; (2) a parameter-efficiency comparison against classical baselines showing that a four-qumode CV-QNN exceeds the validation AUC of a 55-parameter classical network while using 67% fewer trainable parameters; (3) a simplified $\Phi \circ D \circ U_1$ CV-QNN architecture that uses 40–45% fewer trainable parameters than the full Killoran et al. (2019a) layer at every circuit width and is significantly more accurate at four qumodes, where it is the best-performing model overall; (4) the identification and mitigation of barren plateau effects in CV-QNN training through input dimensionality reduction and constrained encoding gate sets, which raise the loss-gradient variance by roughly 58 orders of magnitude; and (5) a discussion of the implications of these results for the broader development of room-temperature, edge-deployable quantum machine learning in healthcare.

The remainder of this paper is organized as follows. Section 2 reviews the relevant background and positions the present work within the existing literature. Section 3 describes the proposed hybrid architecture and training methodology. Section 4 details the dataset and implementation. Section 5 presents the experimental results. Section 6 discusses the findings, including an analysis of barren plateau mitigation strategies. Section 7 concludes the paper and outlines directions for future work.

## 2. Background and Related Work

### 2.1 Early Detection of Oral Cancer

Oral cancer is among the most prevalent malignancies worldwide, particularly in low- and middle-income countries where risk factors such as tobacco use, alcohol consumption, and betel quid chewing are widespread (Warin et al.

2021). Epidemiological studies indicate that the five-year survival rate exceeds 80% when oral cancers are detected at an early, localized stage, but falls below 30% once regional or distant metastasis occurs (Geum et al. 2013). Despite the critical importance of early diagnosis, conventional diagnostic approaches, including biopsy, histopathological analysis, and specialist clinical examination, remain inaccessible in many regions due to high cost, limited availability of trained clinicians, and infrastructural constraints (Uthoff et al. 2018; Warin et al. 2021).

Smartphones, now ubiquitous across both developed and developing regions, present a compelling platform for accessible and scalable screening. Image-based diagnostic tools using smartphone-acquired oral images have demonstrated potential to facilitate early detection and triage in community and primary-care settings, offering a low-cost complement to traditional clinical workflows (Uthoff et al. 2018; Swamikannan et al. 2024). However, mobile-based screening systems must rely on lightweight, energy-efficient, and robust models capable of operating under resource constraints. This requirement motivates the exploration of compact machine learning architectures (Jubair et al. 2022; Swamikannan et al. 2024) and hybrid classical–quantum approaches that emphasize parameter efficiency and deployability (Cong et al. 2019; Mari et al. 2020).

### 2.2 Machine Learning in Medical Imaging

Machine learning (ML) and deep learning, particularly convolutional neural networks (CNNs), have transformed medical image analysis by enabling automated feature extraction and end-to-end learning from raw imaging data (LeCun et al. 2015; Jubair et al. 2022; Khanagar et al. 2023). CNN-based approaches have achieved state-of-the-art performance in a range of medical imaging tasks, including tumor detection, histopathological image classification, and retinal disease screening (Litjens et al. 2017; Ting et al. 2019; Figueroa et al. 2022), by learning hierarchical representations that capture both local and global visual patterns (Kulkarni et al. 2023). Despite their success, conventional CNN architectures often require large annotated datasets (Warin et al. 2021) and substantial computational resources (Sze et al. 2017), which can limit their applicability in low-data and resource-constrained medical settings such as early oral cancer detection (Jubair et al. 2022).

To address deployment constraints in real-world clinical and community environments, significant effort has been devoted to the development of lightweight and edge-optimized neural networks (Jubair et al. 2022). MobileNet, which forms the basis of our approach, employs depthwise separable convolutions to substantially reduce parameter count, memory footprint, and inference latency while maintaining competitive accuracy on standard vision benchmarks (Howard et al. 2017). These properties make it particularly well suited for smartphone-based and point-of-care applications, where energy efficiency, real-time inference, and limited on-device memory are critical considerations (Sze et al. 2017; Jubair et al. 2022).

While lightweight architectures alleviate the compute bottleneck, they do not address the data bottleneck inherent to medical imaging: feature representations extracted by deep CNNs remain high-dimensional, and downstream classifiers trained on small medical datasets are prone to overfitting in this high-dimensional space (Morovati et al. 2023). Dimensionality reduction techniques such as principal component analysis (PCA) have therefore been widely adopted to compress learned feature embeddings, reduce redundancy, and improve generalization while preserving the most informative components (Jolliffe and Cadima 2016). In medical imaging pipelines, PCA has been shown to lower computational cost for downstream stages, mitigate overfitting in small-sample regimes, and provide compact, decorrelated feature inputs suitable for a wide range of subsequent classifiers (Fati et al. 2022; Morovati et al. 2023). Even with both lightweight architectures and dimensionality reduction in place, the underlying classical learning paradigm imposes its own limits on expressivity per trainable parameter, particularly under the strict energy and memory budgets of portable, battery-powered devices (Sze et al. 2017). These limitations have motivated exploration of alternative and complementary computing paradigms, including hybrid classical–quantum learning frameworks, which aim to achieve expressive nonlinear decision boundaries with fewer trainable parameters (Benedetti et al. 2019; Schuld et al. 2020; Kulkarni et al. 2023). Quantum-enhanced learning models, particularly when integrated with classical feature extractors, offer a promising avenue for reducing model complexity while maintaining competitive performance, thereby addressing key bottlenecks in edge-based medical image analysis (Mari et al. 2020; Kulkarni et al. 2023).

### 2.3. Quantum Machine Learning and Hybrid Architectures

**Quantum Machine Learning (QML)** seeks to exploit the principles of quantum mechanics, superposition, entanglement, and interference, to enhance pattern recognition and learning efficiency (Lloyd et al. 2013; Killoran et al. 2019a). The architecture of QML is composed of

- **Data Encoding**
  Classical input features are embedded into quantum states by mapping feature values to the parameters of continuous-variable quantum gates. In practice, displacement gates and, when appropriate, squeezing gates are applied to initialized optical modes (qumodes), transforming classical data into quantum states whose quadrature amplitudes represent the input feature space (Killoran et al. 2019a; Choe and Perkowski 2022).
- **Quantum Circuit (Quantum Neural Network Layers)**
  The encoded quantum states are processed through a parameterized quantum optical circuit composed of interferometers (beam splitters and phase-shift gates), squeezing operations, and displacement gates. The interferometric and squeezing components implement linear transformations equivalent to matrix multiplication, while displacement gates perform affine bias additions. Nonlinear optical elements, such as Kerr gates, introduce non-Gaussianity and act as quantum analogues of activation functions. Multiple stacked layers of these operations constitute a deep continuous-variable quantum neural network (Killoran et al. 2019a; Choe and Perkowski 2022).
- **Measurement and Output**
  After circuit execution, the quantum states of the qumodes are measured using photon-number-resolving detectors or homodyne detection. Measurement outcomes are mapped to classical values that serve as the network outputs or class probabilities. These outputs are used to compute a loss function, and circuit parameters are updated through a hybrid training loop using classical backpropagation (Killoran et al. 2019a; Choe and Perkowski 2022).

While full-scale fault-tolerant quantum computers remain distant, **hybrid classical–quantum models** have emerged as promising near-term strategies (Benedetti et al. 2019; Bharti et al. 2022). These architectures integrate classical preprocessing (e.g., CNNs, PCA) with parameterized quantum circuits that act as nonlinear feature maps or classifiers (Schuld and Killoran 2019; Benedetti et al. 2019; Mari et al. 2020).

Hybrid QML has shown promise in low-resource settings such as small biomedical datasets, where **variational quantum circuits (VQCs)** capture complex correlations with fewer trainable parameters than deep neural networks (Schuld et al. 2020; Kulkarni et al. 2023). However, most existing implementations rely on **qubit-based systems**, which face practical limitations, namely, the requirement for cryogenic operation, restricted qubit coherence times, and complex calibration (Havlíček et al. 2019; Bharti et al. 2022). These constraints limit their immediate applicability to portable or embedded systems.

### 2.4 Continuous-Variable Quantum Computing

The continuous-variable (CV) model of quantum computing, originally introduced by Lloyd and Braunstein (1999), provides a natural framework for quantum information processing in optical systems by encoding information in the continuous degrees of freedom of bosonic modes. Unlike qubit-based architectures that rely on discrete two-level systems and cryogenic operation, CV quantum computing leverages mature photonic technologies operating at room temperature, making it particularly attractive for near-term, scalable implementations. These properties position CV quantum computing as a compelling platform for hybrid machine learning architectures that emphasize parameter efficiency, differentiability, and hardware practicality.

#### 2.4.1 Photonic Quantum Processing Units

Photonic quantum processing units (QPUs) constitute a leading hardware platform for continuous-variable (CV) quantum computing, leveraging the quantum properties of light to perform information processing (Killoran et al. 2019a). In the CV photonic paradigm, quantum information is encoded in bosonic modes (qumodes), typically corresponding to the quadrature amplitudes of optical fields. Early experimental prototypes demonstrated that programmable quantum optical processors with a small number of qumodes, on the order of eight modes, can be realized using integrated photonic components operating at room temperature, providing a concrete pathway toward scalable quantum optical computation (Arrazola et al. 2021).

A defining advantage of photonic QPUs is their ability to operate without cryogenic cooling. Unlike superconducting or spin-based qubit systems that require millikelvin temperatures to maintain quantum coherence, photonic platforms exploit optical degrees of freedom that remain quantum at ambient conditions. Nonclassical light states are generated through optical squeezing, in which a classical laser source is transformed into a quantum state by reducing uncertainty in one quadrature below the shot-noise limit at the expense of increased uncertainty in the conjugate quadrature. These squeezed states form the fundamental quantum resource for CV computation and enable entanglement across multiple optical modes.

Quantum information processing in photonic QPUs is realized through sequences of quantum optical operations implemented as reconfigurable optical circuits. Linear optical components such as beam splitters and phase shifters, combined with squeezing operations and, in some cases, non-Gaussian elements, constitute programmable quantum circuits capable of implementing arbitrary Gaussian transformations (Killoran et al. 2019b). Measurement is typically performed using homodyne detection or photon-number-resolving detectors, which convert optical signals into classical outcomes suitable for hybrid classical–quantum feedback and optimization.
From a systems perspective, photonic QPUs offer strong potential for scalability and integration. Advances in silicon photonics and integrated optical circuits allow multiple optical components to be fabricated on a single chip, reducing footprint and improving stability (Arrazola et al. 2021). Furthermore, photonic platforms naturally interface with classical optical communication infrastructure, suggesting a pathway toward tight integration with edge devices, data centers, and future quantum-enhanced sensing and computing systems.

In contrast, qubit-based quantum processors, particularly those based on superconducting circuits, face inherent challenges for deployment beyond laboratory environments. Cryogenic cooling requirements, sensitivity to environmental noise, and limited qubit coherence times impose significant constraints on portability, power consumption, and system complexity (Krantz et al. 2019; Bharti et al. 2022). These factors currently hinder the feasibility of qubit-based QPUs for embedded or mobile applications (Preskill 2018). Photonic CV architectures, by operating at room temperature and relying on mature optical technologies, present a compelling alternative for near-term hybrid machine learning systems that prioritize deployability and energy efficiency over fault-tolerant universal computation (Killoran et al. 2019a; Choe and Perkowski 2022).

### 2.4.2 Controlled Computational Space in Continuous-Variable Quantum Computing

A key advantage of continuous-variable (CV) quantum computing lies in its ability to access and regulate a high-dimensional computational space using physically meaningful parameters. While the Hilbert space of a single bosonic mode is, in principle, infinite dimensional, practical CV quantum systems operate within an effective finite subspace determined by energy constraints and numerical truncation (Killoran et al. 2019b). This controlled restriction enables CV quantum models to exploit high-dimensional representations while remaining compatible with near-term photonic hardware.

For a CV system consisting of M qumodes with a Fock-state cutoff dimension $d$, the effective Hilbert space dimension scales as (Choe and Perkowski 2022)

$$\dim(\mathcal{H}) = d^{\mathrm{M}}$$

This scaling provides a flexible mechanism to tune the representational capacity of the quantum model by adjusting either the number of qumodes or the cutoff dimension. For instance, a two-qumode system with a cutoff of $d = 10$ spans a 100-dimensional Hilbert space, whereas an equivalent dimensionality in a qubit-based architecture would require at least seven qubits. CV systems can therefore realize equivalent representational capacity with fewer physical components, an attractive property for the resource-constrained settings characteristic of edge deployment.

From an architectural perspective, the ability to regulate computational complexity through cutoff dimension and qumode count aligns naturally with photonic quantum processing units operating at room temperature (Arrazola et al. 2021; Choe and Perkowski 2022). By leveraging high-dimensional continuous degrees of freedom without a proportional increase in physical resources, CV quantum computing offers a scalable and hardware-efficient foundation for quantum-enhanced machine learning models, forming a natural bridge between photonic QPUs and continuous-variable quantum neural networks.

### 2.4.3 Continuous-Variable Quantum Neural Networks

Continuous-variable quantum neural networks (CV-QNNs) provide a natural and expressive framework for implementing neural network–like architectures within the photonic quantum computing paradigm. In contrast to qubit-based quantum circuits, which are restricted to unitary operations acting on discrete two-level systems, CV-QNNs operate on bosonic modes and support a richer class of quantum optical transformations that more closely mirror the components of classical neural networks (Killoran et al. 2019a).

A classical neural network layer is typically expressed as

$$L(x) = \phi(Wx + b),$$

where $x$ denotes the input features, $W$ is a weight matrix encoding linear connections between neurons, $b$ represents bias terms, and $\phi(\cdot)$ is a nonlinear activation function. CV-QNN architectures can faithfully reproduce this structure using parameterized quantum optical gates. Linear transformations analogous to matrix multiplication are implemented through interferometric networks composed of beam splitters and phase-shift (rotation) gates, optionally interleaved with squeezing operations. Collectively, these elements realize arbitrary Gaussian transformations, which can be interpreted as quantum analogues of linear layers and are mathematically equivalent to a singular value decomposition (Killoran et al. 2019a).

Bias terms are naturally incorporated through displacement gates, which translate the quadrature amplitudes of each qumode and thus perform affine transformations on the encoded data. Crucially, nonlinearity, an essential ingredient for deep learning, is introduced through nonlinear optical elements such as Kerr gates (Choe and Perkowski 2022). These non-Gaussian operations enable the implementation of activation-like functions that go beyond purely linear transformations, allowing CV-QNNs to express complex decision boundaries. Stacking multiple such layers yields a genuinely deep quantum neural network, distinguished by the presence of both affine transformations and nonlinear activations.

In contrast, qubit-based quantum circuits are fundamentally limited to linear unitary transformations on the Bloch sphere, making it difficult to implement bias terms and nonlinear activation functions in a manner directly analogous to classical neural networks (Choe and Perkowski 2022). This limitation restricts their expressivity for deep learning tasks. CV-QNNs, by leveraging continuous degrees of freedom and nonlinear optical effects inherent to photonic systems, offer a more faithful and hardware-native realization of neural network computation, particularly well-suited for hybrid classical–quantum learning in practical, near-term settings.

## 2.5. Scope of the Present Work

This work builds on the emerging literature in hybrid CV-classical architectures and addresses both methodological and practical gaps in the deployment of quantum-enhanced models for medical image classification. The present study focuses on two key aspects: the design of a simplified and edge-efficient CV-QNN architecture suited to near-term photonic hardware, and the identification and mitigation of training instabilities arising from the barren plateau phenomenon in continuous-variable quantum circuits.

### 2.5.1 Simplified CV-QNN Architecture

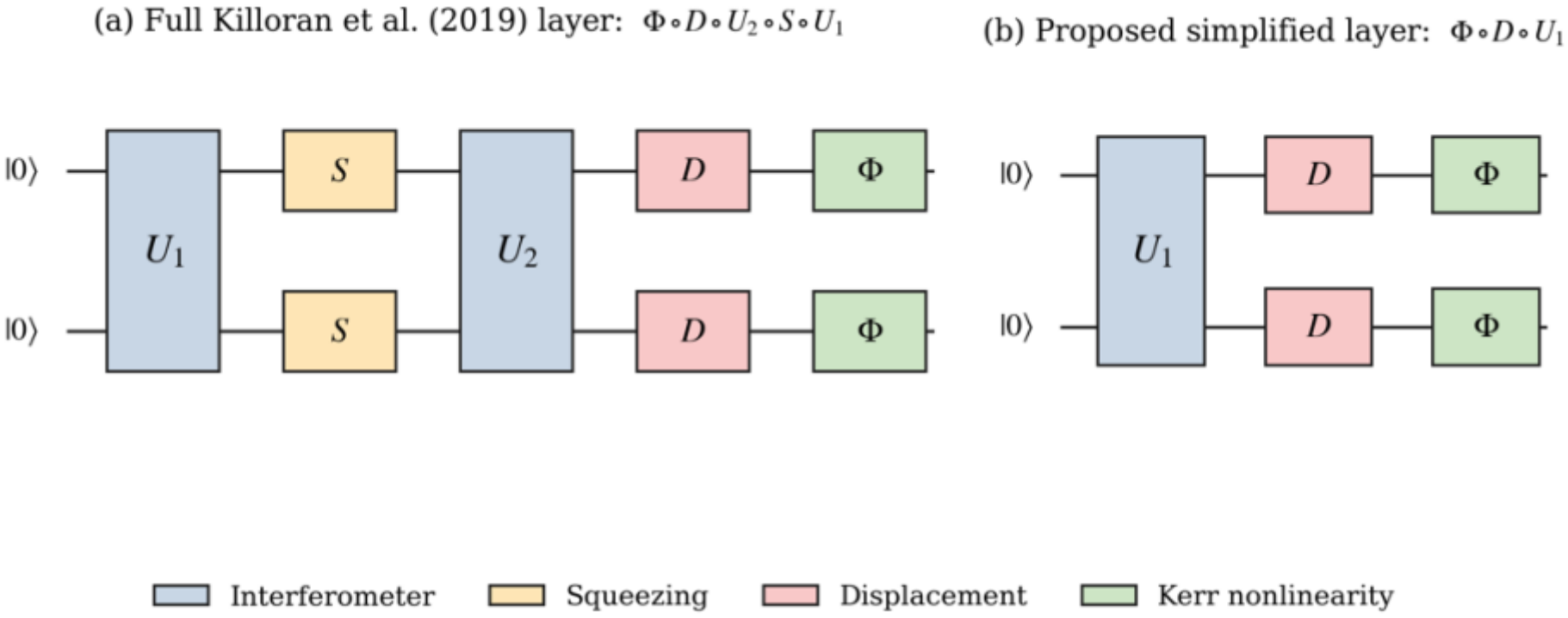

**Fig. 1** Comparison of CV-QNN layer architectures. (a) The full Killoran et al. (2019a) layer implements a generalized affine transformation through an SVD-like decomposition $\Phi \circ D \circ U_2 \circ S \circ U_1$, combining two interferometric blocks, squeezing, displacement, and Kerr nonlinearity. (b) The proposed simplified layer $\Phi \circ D \circ U_1$ retains the displacement bias and Kerr nonlinearity but restricts the linear transformation to a single interferometric block, reducing trainable parameters by over 40% while preserving expressive capacity for hybrid learning with a classical feature extractor. Input modes are initialized in the vacuum state $|0\rangle$

The full CV-QNN architecture introduced by Killoran et al. (2019a) implements a general affine transformation followed by nonlinearity using an SVD-like decomposition expressed as $\Phi \circ D \circ U_2 \circ S \circ U_1$. In this architecture, the interferometric unitaries $U_1$ and $U_2$, composed of beam splitters and rotation gates, perform mode mixing, the squeezing operation S encodes singular value scaling, the displacement operation D introduces learnable bias terms, and the nonlinear activation Φ is realized through Kerr or measurement-induced effects (Fig. 1a). This configuration provides a direct quantum analogue of a fully connected neural network layer with bias and nonlinear activation.

The present work, shown in Fig. 1, adopts a simplified quantum model expressed as $\Phi \circ D \circ U_1$, in which the linear transformation is restricted to a single unitary interferometric layer without explicit squeezing. In this case, $U_1$ implements an orthogonal transformation that preserves feature norms, while D and Φ retain bias addition and nonlinearity, respectively (Fig. 1b). Although this architecture does not realize a full SVD-based linear mapping, it yields a parameter-efficient and implicitly regularized quantum layer with reduced circuit depth and training complexity. When combined with nonlinear activation and, where applicable, multiple stacked layers, the simplified CV-QNN retains sufficient expressive power for hybrid learning, particularly in settings where the quantum circuit operates on high-level features already extracted by a classical network (Mari et al. 2020).

Preliminary single-seed experiments suggested that the simplified CV-QNN could match or exceed the full Killoran et al. architecture while using substantially fewer trainable parameters. The full multi-seed evaluation reported in Section 5, however, reveals a more nuanced, width-dependent picture: at two qumodes the full Killoran layer in fact holds a small but statistically significant advantage (Wilcoxon $p = 0.031$), whereas at four qumodes the simplified layer becomes significantly better while using 44% fewer parameters and yields the best-performing model in the entire study. The reduced gate set also acts as an implicit regularizer that helps keep the circuit trainable on the small PCA-reduced feature set. We therefore retain both the simplified and full layers as experimental conditions throughout and compare them directly in Section 5.5, rather than committing to a single architecture a priori.

### 2.5.2 The Barren Plateau Problem

A further practical challenge encountered in this work is the barren plateau phenomenon, a well-documented training pathology in variational quantum circuits in which the gradients of the cost function vanish exponentially with the size of the quantum system (McClean et al. 2018). This effect arises because randomly initialized parameterized quantum circuits tend to produce states that are nearly uniformly distributed over the accessible Hilbert space, rendering the optimization landscape exponentially flat across almost all parameter configurations. Several factors are known to intensify this problem: deep or highly expressive circuits that approximate unitary 2-designs suppress gradient magnitudes as the system scales (McClean et al. 2018), global cost functions that aggregate measurement outcomes across all modes cause exponential concentration even in shallow circuits (Cerezo et al. 2021), and hardware noise can independently flatten the loss landscape by driving the quantum state toward the maximally mixed state (Wang et al. 2021). Unstructured parameter initialization further compounds these effects by placing the circuit in regions of the landscape where no useful gradient signal exists (Grant et al. 2019).

Although these foundational results were established for discrete-variable (qubit-based) circuits, the barren plateau phenomenon persists, and takes a structurally distinct form, in continuous-variable (CV) systems. Because the underlying CV Hilbert space is infinite-dimensional, standard analytical tools such as unitary t-designs do not directly apply, and dedicated theoretical treatments have been required to characterize trainability in this regime (Zhang and Zhuang 2025). In particular, Zhang and Zhuang showed that in bosonic CV variational circuits the variance of the cost gradient decays exponentially with the number of optical modes but only polynomially with the per-mode circuit energy, identifying an energy-dependent barren plateau without a direct discrete-variable analogue. A practical consequence is that when high-dimensional classical features are directly encoded using a fully expressive set of CV optical gates, the resulting circuits become excessively deep, parameter-heavy, and energy-rich, rapidly entering the regime where gradient magnitudes decay to negligible levels. The choice of encoding strategy, including which gates are used, their ordering, and the dimensionality of the input being encoded, therefore critically determines whether a CV-QNN remains trainable or collapses into a barren plateau. The specific strategies employed in this work to address these training instabilities, namely reducing the input dimensionality prior to quantum encoding, restricting the encoding gate set to displacement and squeezing operations, and adopting a displacement-before-squeezing gate ordering, are examined in detail in Section 6.3 alongside their observed impact on convergence behavior.

Taken together, the simplified architecture and the barren plateau mitigation strategies adopted in this study demonstrate that continuous-variable hybrid quantum models offer a feasible route toward low-power, compact, and deployable quantum inference. This aligns with the broader vision of the Noisy Intermediate-Scale Quantum (NISQ) era (Preskill 2018), a period defined by the use of moderate-scale, error-prone quantum devices that operate without full fault tolerance and rely on hybrid classical-quantum strategies to extract practical computational value from currently available hardware.

## 3. Methods

This study employs a **hybrid classical–quantum classification framework** for automated detection of oral cancer from smartphone-acquired images. The proposed architecture, illustrated in **Fig. 2**, integrates an edge-efficient classical preprocessing pipeline with a continuous-variable quantum neural network (CV-QNN) classifier.
The hybrid model consists of two main components: a **classical preprocessing stage** and a **quantum neural network stage**. The classical stage transforms input RGB images of size $224 \times 224$ pixels into a compact, low-dimensional feature representation, which is subsequently processed by the quantum circuit. Specifically, the classical preprocessing pipeline comprises:

- A convolutional neural network (CNN) based on **MobileNet-V1 augmented with squeeze-and-excitation (SE) blocks**, designed for efficient feature extraction on resource-constrained devices.
- **Principal component analysis (PCA)** for dimensionality reduction, compressing high-dimensional CNN features into a 16-dimensional vector suitable for quantum processing.

The quantum component of the hybrid classifier consists of:

- **Quantum state preparation (data encoding)**, where the 16-dimensional classical feature vector is embedded into a continuous-variable quantum state.
- A **continuous-variable quantum neural network (CV-QNN)** composed of parameterized quantum optical gates that implement a learnable nonlinear classifier.
- **Measurement**, performed using photon-counting or homodyne detection to convert the final quantum state into classical outputs.

This modular design enables the classical front-end to perform computationally intensive feature extraction, while the quantum back-end focuses on learning expressive decision boundaries using a compact, low-parameter model.

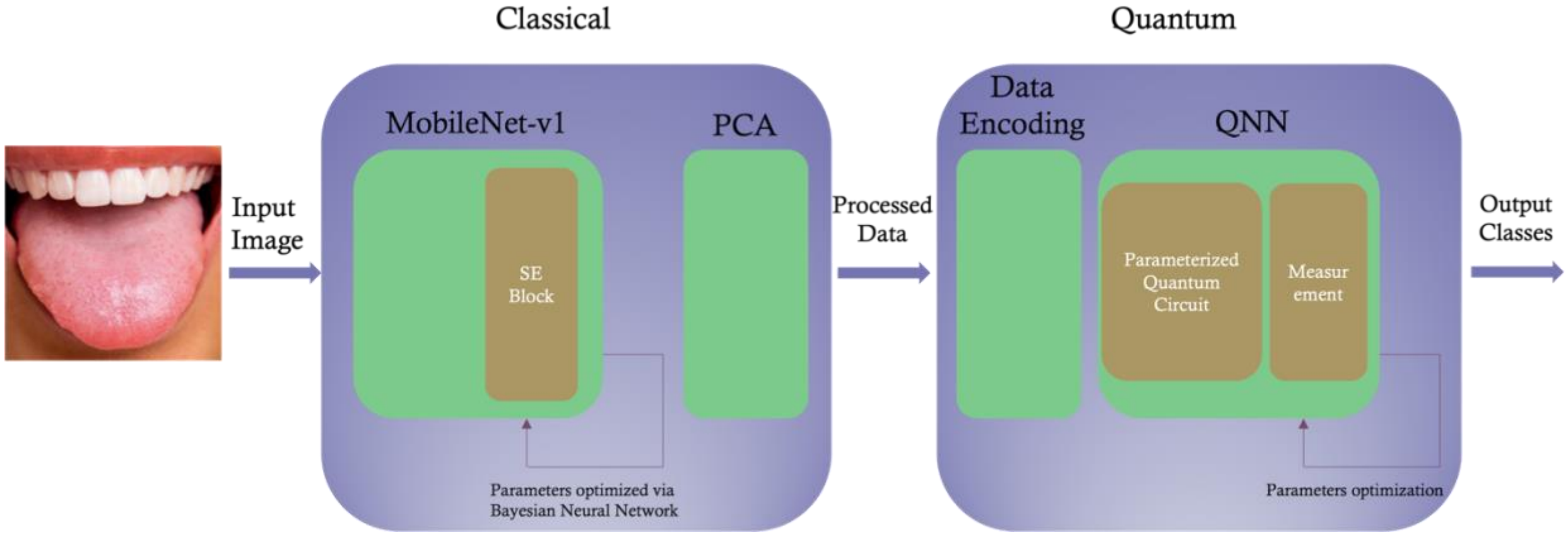

**Fig. 2** Data Flow Diagram of the Proposed Hybrid Classical-Quantum Architecture.

### 3.1 Classical Layer

The classical part of the network is composed of CNN implemented with MobileNet-V1 and Principal Component Analysis. Each input image, sized 224 × 224 pixels with three color channels, is passed through MobileNet-V1 to extract a 1024-dimensional feature vector. PCA is then applied to reduce this vector to the 16 most informative components. These features are converted into a quantum state using a data encoding circuit, then a Quantum Neural Network learns a generalizable pattern to provide the prediction.

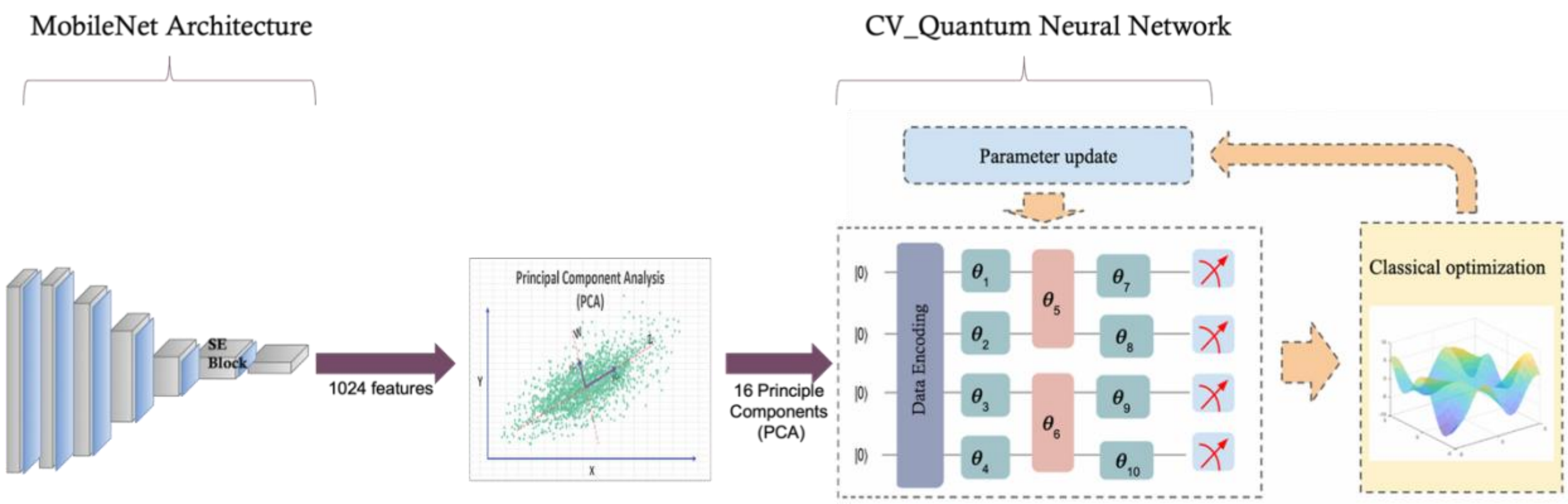

**Fig. 3** Architecture diagram of our Hybrid Classical and Continuous Variable Quantum Neural Network (CV-QNN).

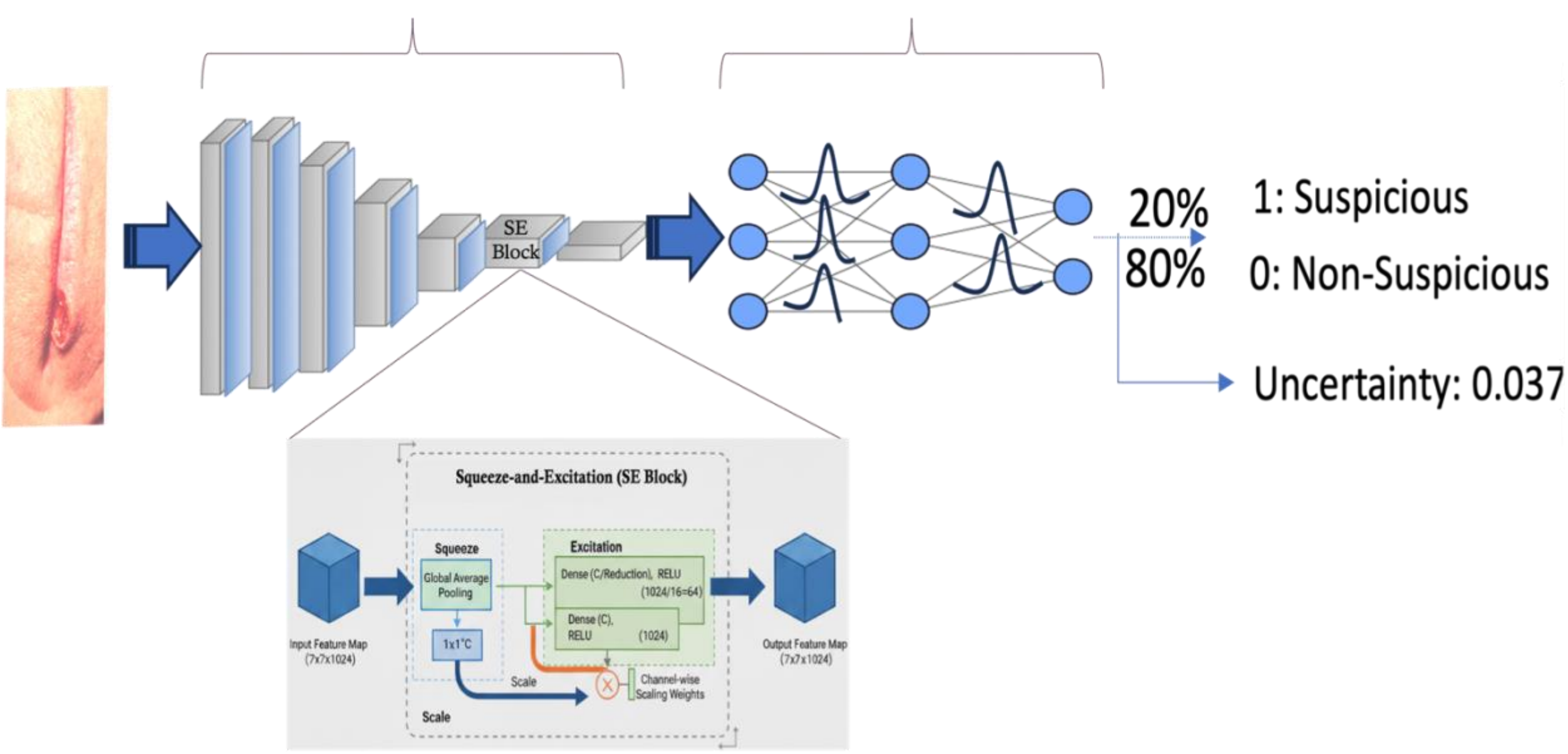

**Fig. 4** Architecture diagram of our Convolutional and Bayesian Neural Network training to learn useful feature representations from smartphone-captured oral cancer images.

### 3.1.1 Convolutional Neural Network (CNN) Layer

The feature extraction pipeline, which serves as the classical front-end of the proposed hybrid architecture, is designed to efficiently transform high-dimensional image data into a compact and discriminative feature representation. As schematically illustrated in **Fig. 3**, this process is initialized using a transfer-learned **MobileNet-V1** backbone, selected for its suitability for deployment on resource-constrained edge devices.

MobileNet-V1 achieves high computational efficiency through the use of **depthwise separable convolutions**, which decompose standard convolutional operations into two specialized steps: (i) a **depthwise convolution**, which applies a single spatial filter independently to each input channel, and (ii) a **pointwise convolution**, which employs a $1 \times 1$ kernel to linearly combine information across channels (Howard et al. 2017).

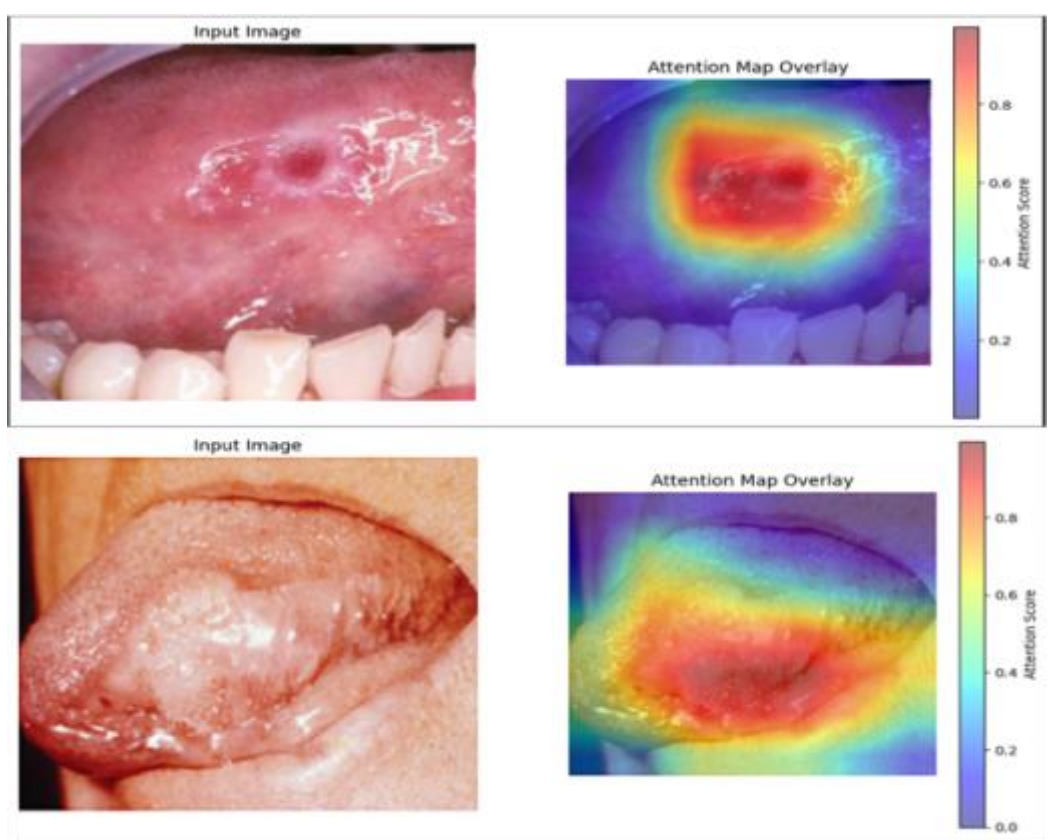

**Fig. 5** The figure displays two examples of input images (left) and their corresponding attention map overlays (right), confirming that the Squeeze-and-Excitation (SE) enhanced MobileNet backbone successfully localizes the oral lesion (high Attention Score) as the most discriminative feature for classification.

To further enhance feature discriminability, a **squeeze-and-excitation (SE) block** was integrated into the final convolutional layer of the network, as shown in **Fig. 4**. The SE mechanism introduces channel-wise attention by dynamically recalibrating feature responses (Hu et al. 2018). Specifically, the SE block operates on the final $7 \times 7 \times 1024$ feature map through two stages. First, a **squeeze** operation applies global average pooling to aggregate spatial information into a $1 \times 1 \times 1024$ channel descriptor. This is followed by an **excitation** operation, in which two fully connected layers process the descriptor to generate a set of learned scale factors ($1 \times 1 \times 1024$) that selectively emphasize informative channels while suppressing less relevant ones (Hu et al. 2018).

The feature refinement introduced by the SE block is critical for ensuring that the extracted representation remains both compact and information rich, as shown in Fig. 5. This property enables subsequent dimensionality reduction via principal component analysis (PCA) to preserve over 95% of the original feature variance while producing a substantially lower-dimensional embedding. The MobileNet-V1 backbone was initially trained using a **Bayesian neural network (BNN) classifier head** with variational inference on the oral cancer dataset as described in our last study Sonawane et al. (2025). The learned parameters from this classical pipeline were subsequently transferred to the hybrid classical–quantum model, enabling stable and discriminative feature extraction for downstream quantum processing.

### 3.1.2 Principal Component Analysis (PCA) Layer

Following feature extraction, **principal component analysis (PCA)** was applied to the resulting 1024-dimensional feature vectors to achieve essential dimensionality reduction. PCA was employed to mitigate overfitting, reduce computational overhead, and produce a compact representation compatible with quantum processing constraints (Jolliffe and Cadima 2016; Benedetti et al. 2019). A total of **16 principal components** were retained, capturing **95.61% of the total variance**, exceeding the 70 to 90% cumulative-variance cut-off points commonly reported in the PCA literature (Jolliffe and Cadima 2016) and adopted in prior medical imaging pipeline (Cangelosi and Goriely 2007).

The resulting standardized $N \times 16$ feature matrix served as the final classical output of the preprocessing pipeline. These 16 features were subsequently encoded into the continuous-variable quantum neural network (CV-QNN), where they were distributed across **two and four qumodes** at the input stage prior to the application of the variational quantum layer. Importantly, this reduced-dimensional feature set was used consistently as input to both the proposed CV-QNN and the baseline classical neural network, ensuring a controlled and fair comparison of classification performance.

## 3.2 Continuous-Variable Quantum Neural Network Layer

### 3.2.1 Data Encoding (Quantum State Preparation)

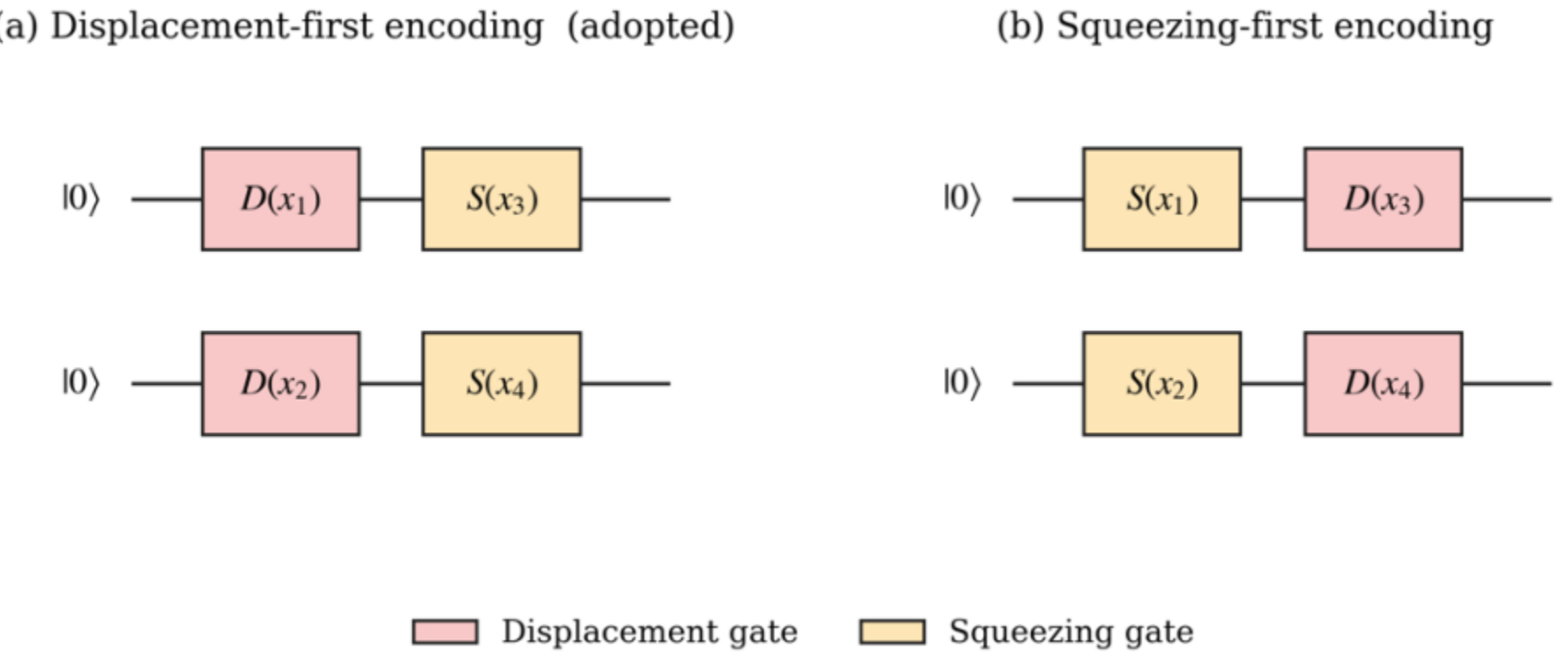

**Fig. 6** The Data encoding circuits are compared in this work. (a) Displacement-first ordering: vacuum modes are first displaced by feature-dependent amplitudes, then squeezed. (b) Squeezing-first ordering: vacuum modes are first squeezed, then displaced. The displacement-first ordering is adopted in the final model, consistent with the encoding strategy discussed in Section 6.3.

For classical data to be processed by a quantum circuit, it must first be encoded into a quantum state, a procedure commonly referred to as **quantum state preparation**. In this process, classical data represented as entries of a feature vector are mapped to the parameters of parameterized quantum gates, thereby embedding classical information into the quantum domain.

In the proposed architecture, classical features are encoded using **continuous-variable quantum optical gates**, specifically:

- **Displacement gates**, which shift the quadrature amplitudes of optical modes and directly encode feature values, and
- **Squeezing gates**, which modify the quantum uncertainty distribution of the quadratures and enhance representational expressivity.

The two possible orderings of these encoding gates are illustrated in Fig. 6. In the displacement-first scheme (Fig. 6a), each qumode is first displaced from the vacuum by an amount proportional to a feature value, and the squeezing operation subsequently deforms the resulting coherent state. In the squeezing-first scheme (Fig. 6b), the same gate types are applied in the reverse order. The displacement-first ordering is adopted here, and the trainability considerations that motivate the encoding strategy are discussed in Sect. 6.3.

Through these operations, the classical feature vector is transformed into a multi-qumode quantum state whose quadrature amplitudes carry the encoded information. The resulting quantum state then serves as the input to the continuous-variable quantum neural network (CV-QNN) for subsequent quantum processing.

### 3.2.2 Quantum Neural Network Circuit

The quantum circuit consists of a sequence of parameterized quantum gates whose optimal parameters are learned through a hybrid training process. This circuit forms the core of the quantum neural network and is followed by a measurement stage that maps quantum states back to classical values. The measured outputs are compared against target labels, and gradients computed via classical backpropagation are used to iteratively update the circuit parameters.

In this study, several architectural variants of the CV-QNN were explored to evaluate the impact of quantum model capacity and expressivity. The experimental design considered the following parameters:

- **Number of qumodes**
  Each qumode is represented by a quantum harmonic oscillator and serves as a computational wire, analogous to a qubit in discrete-variable systems. While photonic platforms such as Xanadu's X8 processor provide up to eight qumodes, only a subset is required for a given experiment. In this work, models utilizing **2 and 4 qumodes** were evaluated.

- **Cutoff dimension**
  Unlike qubit-based systems, where the computational basis is fixed to two states, continuous-variable systems allow each qumode to span multiple Fock states $|0\rangle, |1\rangle, \ldots, |n\rangle$. The **cutoff dimension** defines the maximum number of basis states retained in simulation and thus determines the dimensionality of the Hilbert space. Experiments were conducted with cutoff dimensions of **2, 5, and 10**. For example, a cutoff dimension of 5 corresponds to a five-dimensional computational space per qumode.
- **Number of quantum neural network layers**
  A single CV-QNN layer consists of an ordered sequence of beam splitters and rotation gates (implementing unitary linear transformations), displacement gates (bias addition), and Kerr gates (nonlinear activation). Repeating this sequence yields a deeper quantum neural network. In this study, architectures with **one and two CV-QNN layers** were investigated.

## 4. Dataset and Implementation

### 4.1 Dataset

This study leveraged two oral cancer image datasets to evaluate the proposed parameter-efficient hybrid classical–quantum classifier: (i) a proprietary dataset collected by practicing dentists in India (Mani and Narayana 2024), and (ii) a publicly available dataset from Kaggle (Barot and Suthar 2020). Together, these datasets enabled both in-distribution training and validation as well as out-of-distribution generalization assessment.

The primary dataset consisted of **841 smartphone-acquired color images**, including **290 suspicious** and **551 nonsuspicious** cases. Images were captured under real-world clinical conditions using consumer smartphones and exhibited substantial variability in resolution, illumination, and background. In several instances, extraneous objects such as gloves and dental instruments introduced visual noise.

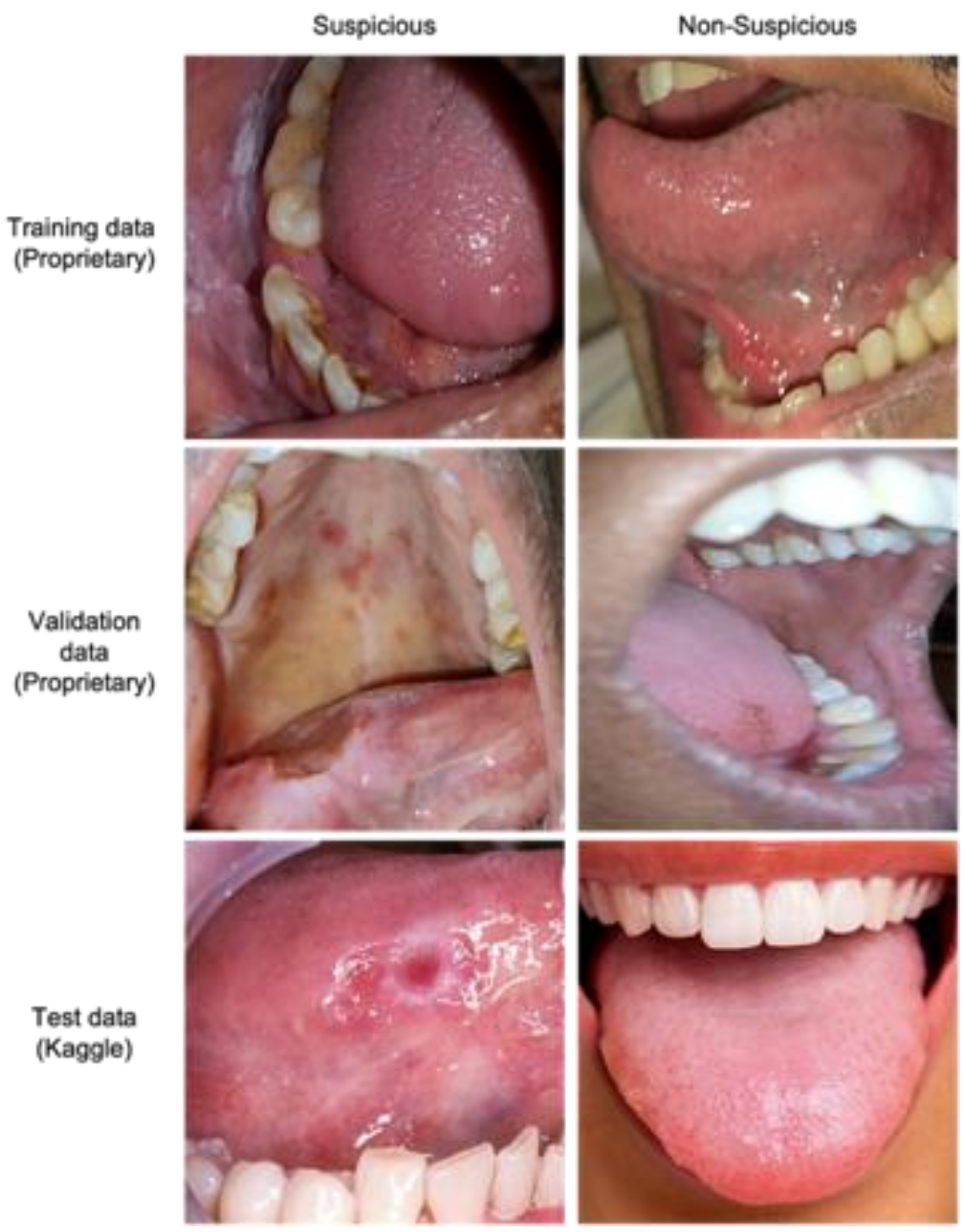

**Fig. 7** Sample images from proprietary dataset and Kaggle dataset.

To mitigate these effects and ensure clinically meaningful feature extraction, all images were **manually cropped** to isolate relevant intraoral regions. This preprocessing step reduced background artifacts while preserving diagnostically important visual cues.

From the proprietary dataset, a balanced subset of **100 images** (**50 suspicious** and **50 nonsuspicious**) was randomly selected to serve as a validation set. This validation set was:

- Completely held out from augmentation steps applied to the training data
- Used exclusively for performance monitoring and hyperparameter tuning during training

This separation ensured an unbiased evaluation of model generalization throughout development.

Given the limited size and high variability of the proprietary dataset, data augmentation was applied to the training set to improve generalization and reduce overfitting. Augmentation strategies were selected to reflect variations commonly observed in smartphone-acquired oral images and included:

- Rotation
- Shearing
- Scaling

Blank or corrupted images generated during augmentation were identified and removed. To further address class imbalance, the majority class was downsampled to match the minority class. Following augmentation and balancing, a total of **3,026 images** were used to train the classical feature extractor.

The second dataset, obtained from Kaggle, comprised **131 color images**, including **87 suspicious** and **44 nonsuspicious** cases, primarily depicting lesions of the lips and tongue. As the proprietary dataset focused predominantly on intraoral regions, most lip images were excluded.

A curated subset exhibiting strong anatomical and textural similarity to the proprietary dataset was retained and cropped to match the same regions of interest. This subset consisted of **31 images** (**21 suspicious** and **10 nonsuspicious**) and was used exclusively as an **external out-of-distribution test set**.

During model development, a validation set was used in place of a test set to enable iterative performance monitoring and hyperparameter tuning. The curated Kaggle dataset was reserved solely for final evaluation, providing a realistic assessment of the model's generalization capability across datasets, acquisition conditions, and patient populations. Figure 7 displayed few samples from our datasets.

### 4.2 Training Strategy and Implementation

Model training follows a **transfer learning–based hybrid optimization strategy**, rather than sequentially training the classical and quantum components from scratch. A fully classical baseline classifier, comprising the pipeline **MobileNet-V1 → Bayesian Neural Network**, was first trained for oral cancer detection, achieving an accuracy of 100% in the test set and an AUC of 96.49% in validation set. The learned parameters from the MobileNet-V1 stage were then transferred to initialize the hybrid architecture, allowing the quantum neural network to focus exclusively on optimizing its own trainable parameters.

This preprocessing strategy after augmenting with PCA serves two critical purposes. First, it minimizes computational overhead by reducing the dimensionality of the feature space from 1024 to 16. Second, it mitigates the **barren plateau problem**, a well-known challenge in variational quantum circuits (VQCs) in which gradients vanish exponentially with increasing circuit depth or input dimensionality. By constraining the quantum model to operate on a compact feature representation, the resulting CV-QNN remains trainable within realistic quantum hardware and simulation constraints.

All CV-QNN models were implemented using the PennyLane quantum machine learning framework with the Strawberry Fields photonic backend, operating on the strawberryfields.fock device. Input features were normalized using Min–Max scaling to the range [0,1] to ensure stable quantum state preparation within the finite cutoff dimension of the photonic modes. The hybrid optimization loop was implemented in PyTorch, with the quantum circuit defined as a differentiable qnode fully integrated into the PyTorch automatic differentiation pipeline.

## 5. Results

This section reports the empirical evaluation of the continuous-variable quantum neural network (CV-QNN) on the oral-cancer screening task. To obtain trustworthy estimates rather than single fortunate runs, the headline model comparisons (Sections 5.1, 5.5 and 5.6) are each repeated across six independent random seeds (40–45) and summarised as mean ± standard deviation. The architectural ablation studies (Sections 5.2–5.4), which isolate the effect of a single hyper-parameter at a time, are run at a fixed seed (42) so that the only source of variation is the factor under study. Every model is trained for 51 epochs with a batch size of 64; quantum parameters use a learning rate of 0.1, classical parameters a learning rate of 0.01, and gradients are clipped to a maximum norm of 1.0. For each run we record the epoch at which the validation area under the ROC curve (AUC) is highest and report the corresponding metrics. We treat the validation AUC as the primary, threshold-independent figure of merit because it is insensitive to the operating point and is the most stable indicator on the small held-out splits used here. In addition to the default decision threshold of 0.5 we report a calibrated test accuracy, obtained by selecting the decision threshold that maximises balanced accuracy on the validation set and applying it unchanged to the test set.

### 5.1 Effect of the number of qumodes

The number of qumodes fixes the width of the photonic circuit and therefore the dimensionality of the quantum feature space available to the model. Table 1 compares the simplified $\Phi \circ D \circ U_1$ layer at two and four qumodes, all with a single trainable layer and local (single-qumode) measurement, averaged over six seeds. Widening the circuit from two to four qumodes raises the validation AUC from 92.75 ± 0.92% (two qumodes, cutoff 5) to 97.67 ± 0.06% (four qumodes, cutoff 5), an improvement of almost five AUC points for the cost of only ten additional parameters. Crucially, increasing the Fock-space cutoff at two qumodes (from 5 to 10) lifts the AUC by less than one point (to 93.51 ± 0.77%), so the gain from a wider circuit is far larger than the gain from a deeper truncation. The four-qumode model also exhibits the smallest seed-to-seed variance of any configuration (± 0.06% AUC), indicating that the additional width yields not only higher but markedly more stable performance. Figure 8 shows that its training, validation and test curves converge smoothly and remain tightly coupled across epochs, with no sign of divergence or overfitting.

*Table 1 Effect of circuit width on the simplified CV-QNN (single layer, local measurement). Values are mean ± standard deviation over six seeds; the four-qumode model (bold) attains the highest validation AUC of any configuration in this study.*

| Configuration (simplified, L=1, local) | #Params | Train acc (%) | Test acc @0.5 (%) | Val AUC (%) |
|---|---|---|---|---|
| 2 qumodes, cutoff d=5 | 8 | 83.78 ± 16.02 | 77.96 ± 17.61 | 92.75 ± 0.92 |
| 2 qumodes, cutoff d=10 | 8 | 83.81 ± 13.91 | 88.71 ± 18.32 | 93.51 ± 0.77 |
| **4 qumodes, cutoff d=5** | **18** | **92.04 ± 10.64** | **97.85 ± 3.57** | **97.67 ± 0.06** |

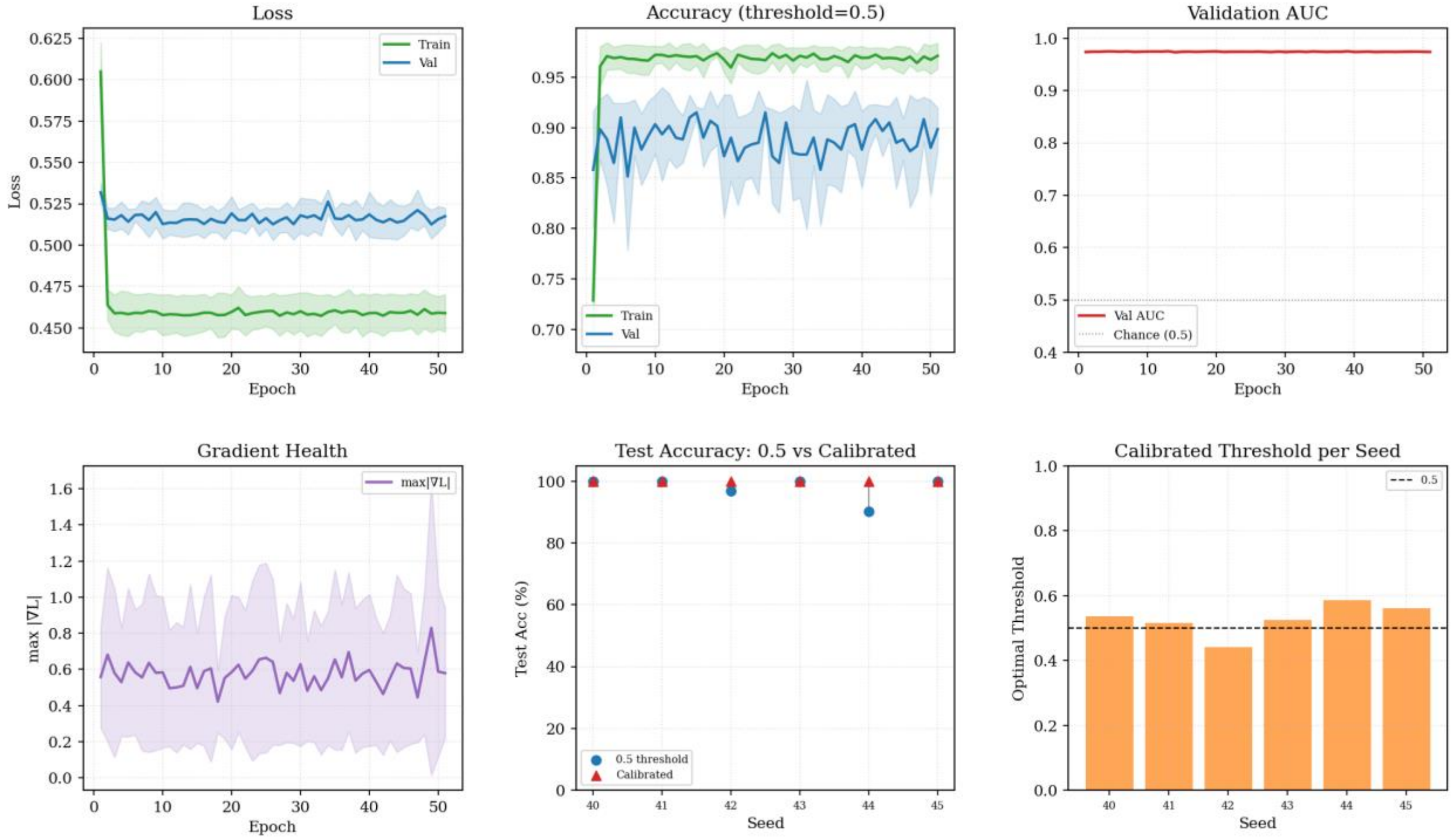

**Fig. 8** Training dynamics of the four-qumode simplified CV-QNN ($\Phi \circ D \circ U_1$, cutoff 5, local measurement). Loss, accuracy and AUC are shown for the training, validation and test splits across 51 epochs.

### 5.2 Effect of the cutoff dimension

The cutoff dimension sets the truncation level of the infinite-dimensional Fock space in which each qumode is represented, and therefore controls how faithfully the simulated photonic state is reproduced. Table 2 sweeps the cutoff for the two-qumode simplified model at the fixed ablation seed. The decisive quantity here is the training accuracy, which reveals whether the optimiser is able to fit the data at all. At cutoffs of 2 and 5 the training accuracy sits at the chance level (≈50%): the Fock truncation is so aggressive that the encoded states cannot separate the classes and the model collapses to a near-constant output, even though the rank-based validation AUC remains deceptively high on the small validation split. A cutoff of 7 begins to fit the training set (79.02%) but still yields an unreliable classifier whose test accuracy falls below chance. Only at a cutoff of 10 does the model converge cleanly, reaching 96.40% training accuracy, 87.00% validation accuracy and a test accuracy of 100% at the default 0.5 threshold. We therefore adopt a cutoff of 10 for the two-qumode experiments. The gradient-level explanation for the failure of small cutoffs is given in Section 6.3.

*Table 2 Effect of the Fock-space cutoff dimension on the two-qumode simplified CV-QNN (single layer, local measurement, seed 42). Training accuracy is the key indicator of convergence; the chosen operating point (d = 10, bold) is the smallest cutoff at which the model fits the training data and produces a thresholdable classifier.*

| Cutoff dimension | Train acc (%) | Val acc (%) | Test acc @0.5 (%) | Val AUC (%) |
|---|---|---|---|---|
| d = 2 | 51.16 | 50.00 | 67.74 | 95.76 |
| d = 5 | 49.74 | 78.00 | 77.42 | 93.80 |
| d = 7 | 79.02 | 51.00 | 45.16 | 91.92 |
| **d = 10** | **96.40** | **87.00** | **100.00** | **93.24** |

### 5.3 Effect of the measurement strategy

We compare two read-out strategies for mapping the photonic state back to a classical decision variable: a local strategy that measures the mean photon number of a single qumode (qumode 0), and a global strategy that measures and combines the photon-number expectations of all qumodes in the circuit. Table 3 reports both at the two-qumode, cutoff-10, single-layer configuration. Local measurement dominates the global alternative on every metric, improving training accuracy by 9.5 points (96.40% versus 86.95%), validation accuracy by 10 points and validation AUC by more than 2 points, while raising the test accuracy to 100%. Reading out a single, well-optimized qumode evidently isolates more of the discriminative structure encoded in the quantum state than aggregating measurements across all modes; this is consistent with the local-versus-global cost-function distinction known to govern trainability in variational circuits, where local observables retain informative gradients while global observables tend toward exponential concentration. Local measurement is therefore used throughout the remaining experiments.

*Table 3 Effect of the measurement strategy on the two-qumode simplified CV-QNN (cutoff 10, single layer, seed 42). The local read-out, which measures a single qumode (qumode 0; top row), outperforms the global read-out, which combines all qumodes, on every metric.*

| Measurement strategy | Train acc (%) | Val acc (%) | Test acc @0.5 (%) | Val AUC (%) |
|---|---|---|---|---|
| Local (qumode 0) | 96.40 | 87.00 | 100.00 | 93.24 |
| Global (all qumodes) | 86.95 | 77.00 | 96.77 | 91.04 |

### 5.4 Effect of the circuit depth

Circuit depth is the number of sequential trainable layers stacked in the variational circuit; each additional layer adds expressive power at a linear cost in parameters. Table 4 compares one and two layers for the two-qumode, cutoff-10 simplified model. A second layer doubles the parameter count from 8 to 16 yet changes the validation AUC by only 0.40 points (93.64% versus 93.24%) and slightly lowers both training and validation accuracy, while the test accuracy stays at 100% in both cases. The returns from added depth are therefore marginal at this width, so we retain a single layer as the default, reserving the extra parameter budget for the wider four-qumode circuit that proved far more effective in Section 5.1.

*Table 4 Effect of circuit depth on the two-qumode simplified CV-QNN (cutoff 10, local measurement, seed 42). Doubling the depth yields only a marginal AUC change for twice the parameters.*

| Layers | #Params | Train acc (%) | Val acc (%) | Test acc @0.5 (%) | Val AUC (%) |
|---|---|---|---|---|---|
| L = 1 | 8 | 96.40 | 87.00 | 100.00 | 93.24 |
| L = 2 | 16 | 94.55 | 85.00 | 100.00 | 93.64 |

### 5.5 Simplified versus full Killoran layer

A central question of this work is whether the parameter-efficient simplified layer $\Phi \circ D \circ U_1$ can replace the full Killoran layer $\Phi \circ D \circ U_2 \circ S \circ U_1$, which interleaves an additional interferometer and squeezing stage. The simplified layer uses 40–45% fewer parameters at every width (8 versus 14 at two qumodes, 18 versus 32 at four qumodes). To test the comparison rigorously we apply a Wilcoxon signed-rank test to the six paired per-seed validation-AUC values for each configuration. The outcome is a clear, width-dependent crossover rather than a uniform winner (Table 5, Figure 9). At two qumodes the full Killoran layer is significantly better, at both cutoff 5 (94.42% versus 92.75%) and cutoff 10 (94.81% versus 93.51%), each with $p = 0.031$ and a maximal rank-biserial effect size of −1.0 (the full layer wins on all six seeds). At four qumodes the ordering reverses: the simplified layer becomes significantly better (97.67% versus 97.26%, $p = 0.031$, effect size +1.0), winning on every seed while using 44% fewer parameters. In other words, the extra gates of the full Killoran layer help when the circuit is narrow and the feature space is small, but at four qumodes they become redundant and the leaner simplified layer both wins outright and does so far more economically.

Figure 10 shows the corresponding two-qumode training curves, where the full layer's additional capacity translates into the modest edge reported here.

*Table 5 Simplified versus full Killoran trainable layer (single layer, local measurement). Cells report validation AUC (mean ± standard deviation over six seeds) with the parameter count in parentheses; p-values are from a Wilcoxon signed-rank test on the paired per-seed values. The better layer depends on circuit width: the full layer wins at two qumodes, the simplified layer at four.*

| Configuration | Simplified $\Phi \circ D \circ U_1$ | Full Killoran $\Phi \circ D \circ U_2 \circ S \circ U_1$ | Wilcoxon p | Better layer |
|---|---|---|---|---|
| 2 qumodes, d=5 | 92.75 ± 0.92 (8 p) | 94.42 ± 0.86 (14 p) | 0.031 | Full Killoran |
| 2 qumodes, d=10 | 93.51 ± 0.77 (8 p) | 94.81 ± 0.49 (14 p) | 0.031 | Full Killoran |
| 4 qumodes, d=5 | 97.67 ± 0.06 (18 p) | 97.26 ± 0.03 (32 p) | 0.031 | Simplified |

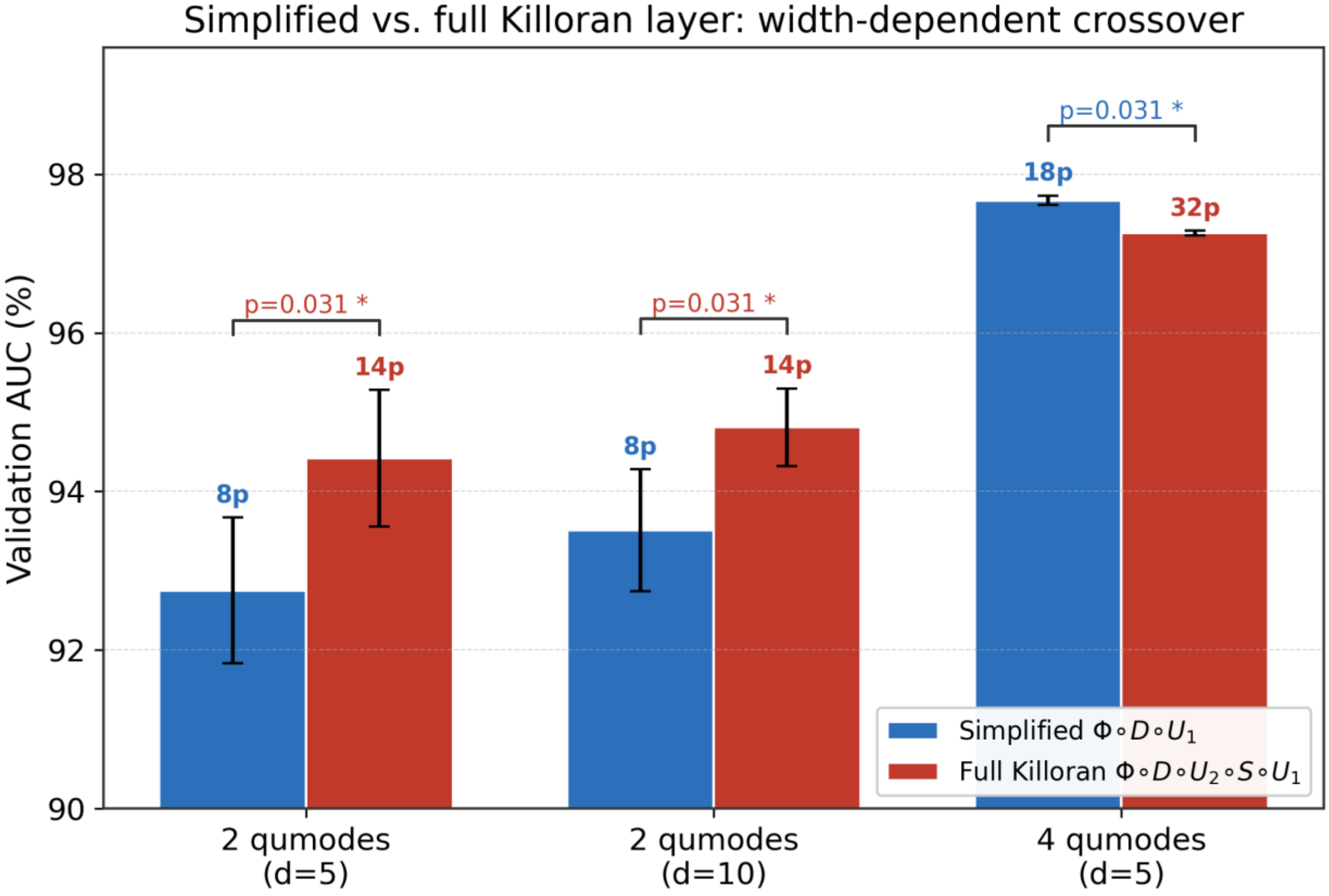

**Fig. 9** Width-dependent crossover between the simplified $\Phi \circ D \circ U_1$ and full Killoran $\Phi \circ D \circ U_2 \circ S \circ U_1$ layers. Bars show validation AUC (mean ± standard deviation over six seeds) with parameter counts annotated; the full layer is significantly better at two qumodes while the simplified layer is significantly better at four qumodes (Wilcoxon signed-rank, $p = 0.031$ in all three comparisons).

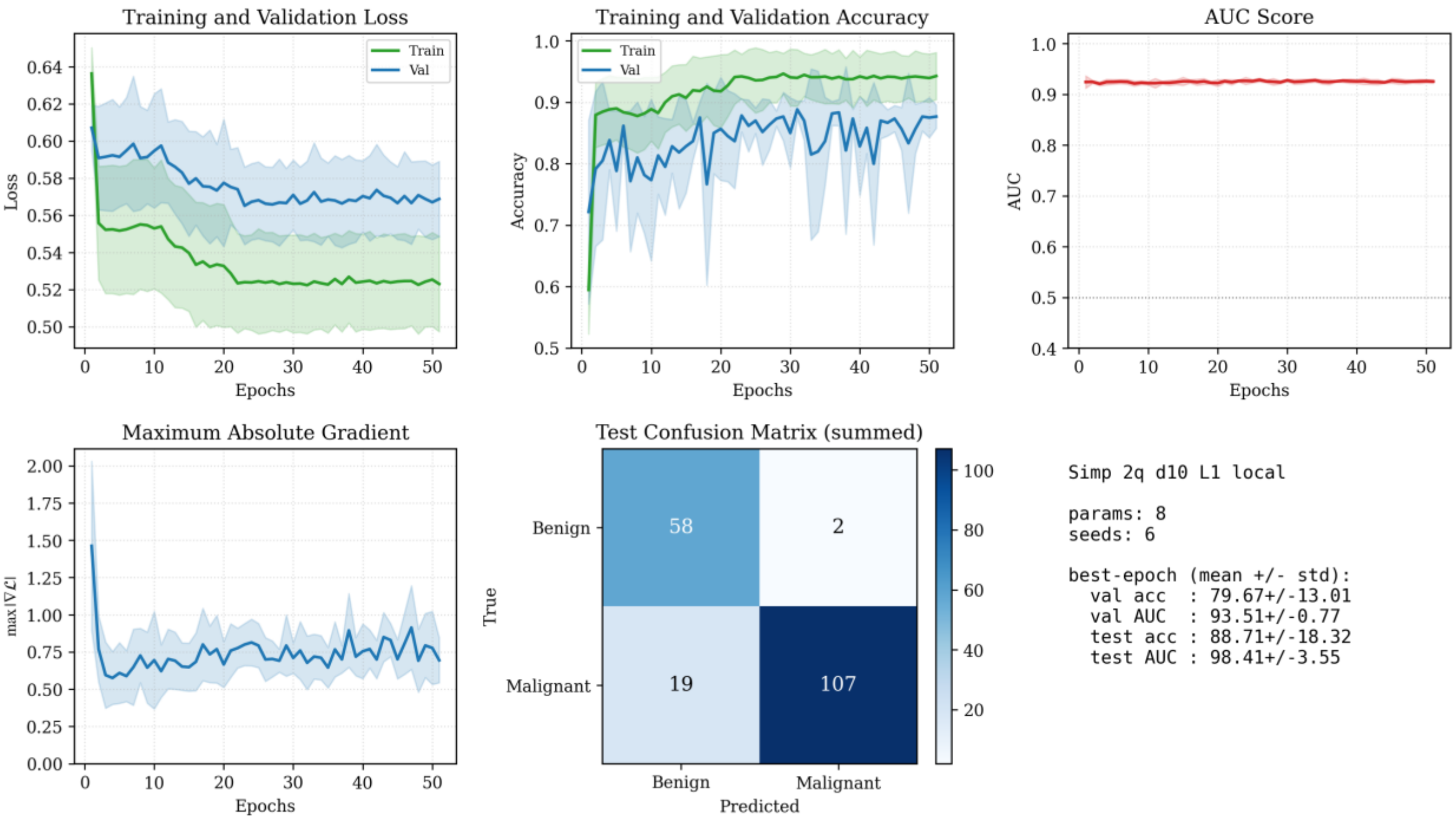

**Fig. 10** Training dynamics of the two-qumode simplified CV-QNN ($\Phi \circ D \circ U_1$, cutoff 10, local measurement) over 51 epochs.

### 5.6 Comparison with classical baselines

To place the quantum models in context we train two classical artificial neural networks (ANNs) on the same 16-dimensional PCA features and the same six seeds. ANN-17 is a single linear unit with 17 parameters, the smallest classical baseline that matches the input dimensionality, and ANN-55 is a compact one-hidden-layer network with 55 parameters. Two findings emerge. First, at two qumodes the quantum models do not beat the classical baselines: the simplified two-qumode model's validation AUC (93.51%) is significantly below both ANN-17 (96.79%) and ANN-55 (97.21%), each with $p = 0.031$. Only by widening to four qumodes does the quantum model overtake the classical networks. Second, and most importantly, the four-qumode simplified CV-QNN attains the highest validation AUC of any model in the study, classical or quantum (97.67 ± 0.06%), edging past ANN-55 (97.21 ± 0.15%) and the full Killoran four-qumode model (97.26 ± 0.03%) while using only 18 trainable parameters, 67% fewer than ANN-55 and 44% fewer than the full Killoran model. All four leading models reach 100% calibrated test accuracy across every seed. Table 6 and Figure 11 summarise this parameter-efficiency frontier, on which the simplified four-qumode model is the Pareto-optimal point: no other model achieves a higher AUC, and no model with a comparable or higher AUC uses fewer parameters. Figure 12 shows the per-seed distribution of the four-qumode and classical models as box plots, confirming that the simplified model's advantage is consistent rather than driven by a single seed.

*Table 6 Headline comparison of the four-qumode CV-QNNs with classical baselines (mean ± standard deviation over six seeds). The simplified four-qumode model (bold) achieves the highest validation AUC of any model while using the fewest parameters of the top performers; all four models reach 100% calibrated test accuracy.*

| Model | #Params | Val AUC (%) | Test acc @0.5 (%) | Test acc calib. (%) |
|---|---|---|---|---|
| ANN-17 (linear, classical) | 17 | 96.79 ± 0.41 | 98.39 ± 2.46 | 100.00 ± 0.00 |
| ANN-55 (1 hidden, classical) | 55 | 97.21 ± 0.15 | 94.62 ± 12.02 | 100.00 ± 0.00 |
| Full Killoran, 4 qumodes | 32 | 97.26 ± 0.03 | 98.92 ± 1.52 | 100.00 ± 0.00 |
| **Simplified, 4 qumodes** | **18** | **97.67 ± 0.06** | **97.85 ± 3.57** | **100.00 ± 0.00** |

Parameter efficiency: validation AUC vs. model size (6 seeds)
Simplified 4q (18p)
best classical baseline
ANN-55 (1 hidden)
Killoran 4q (32p)
ANN-17 (linear)
Killoran 2q (14p)
Simplified 2q (8p)
Simplified CV-QNN ($\Phi \circ D \circ U_1$)
Full Killoran CV-QNN ($\Phi \circ D \circ U_2 \circ S \circ U_1$)
Classical ANN
Validation AUC (%)
Number of trainable parameters

**Fig. 11** Parameter-efficiency frontier: validation AUC against the number of trainable parameters (mean ± standard deviation over six seeds). The simplified four-qumode CV-QNN (circled) is Pareto-optimal, lying above the best classical baseline while using the fewest parameters of the high-AUC models.

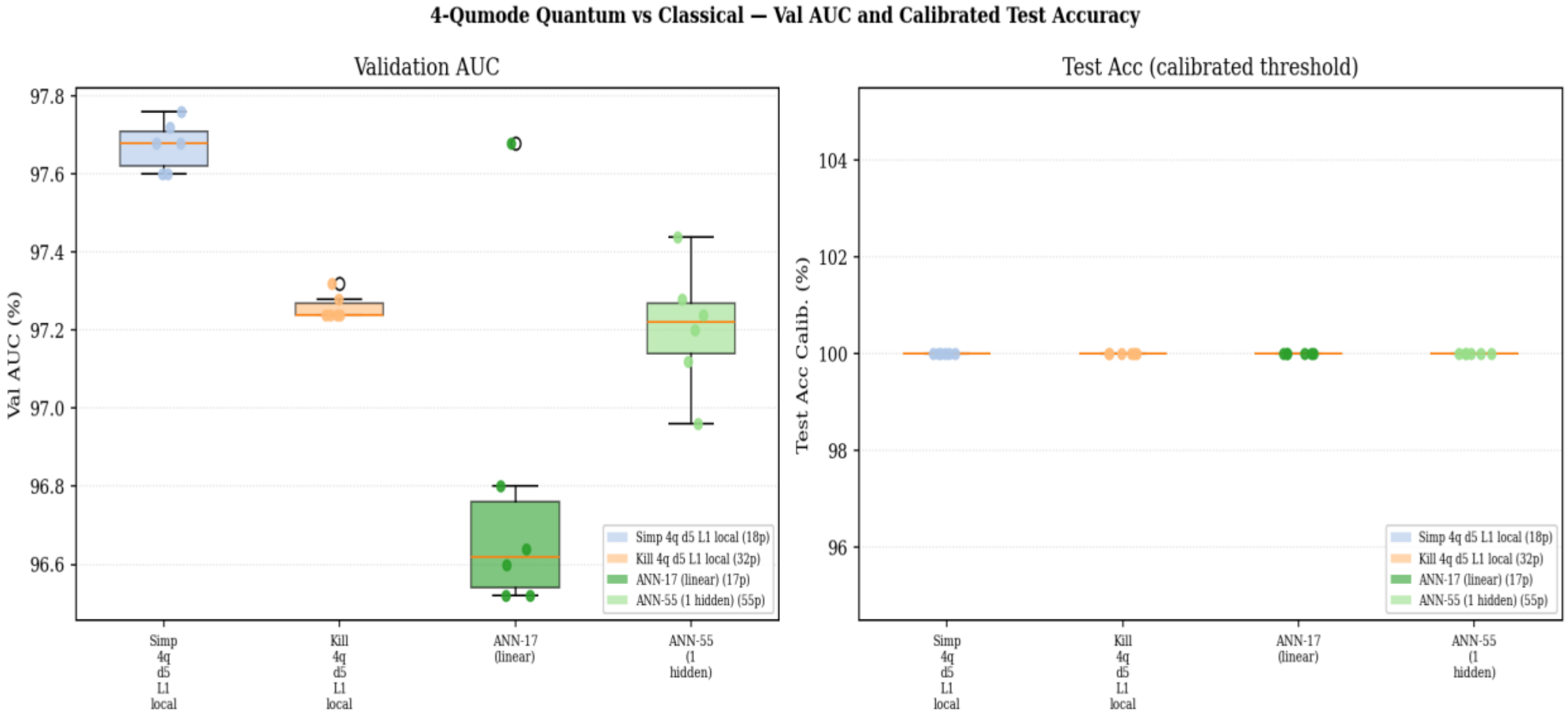

**Fig. 12** Per-seed validation-AUC distributions (box plots over six seeds) for the four-qumode CV-QNNs and the classical baselines, showing that the simplified model's advantage is consistent across seeds.

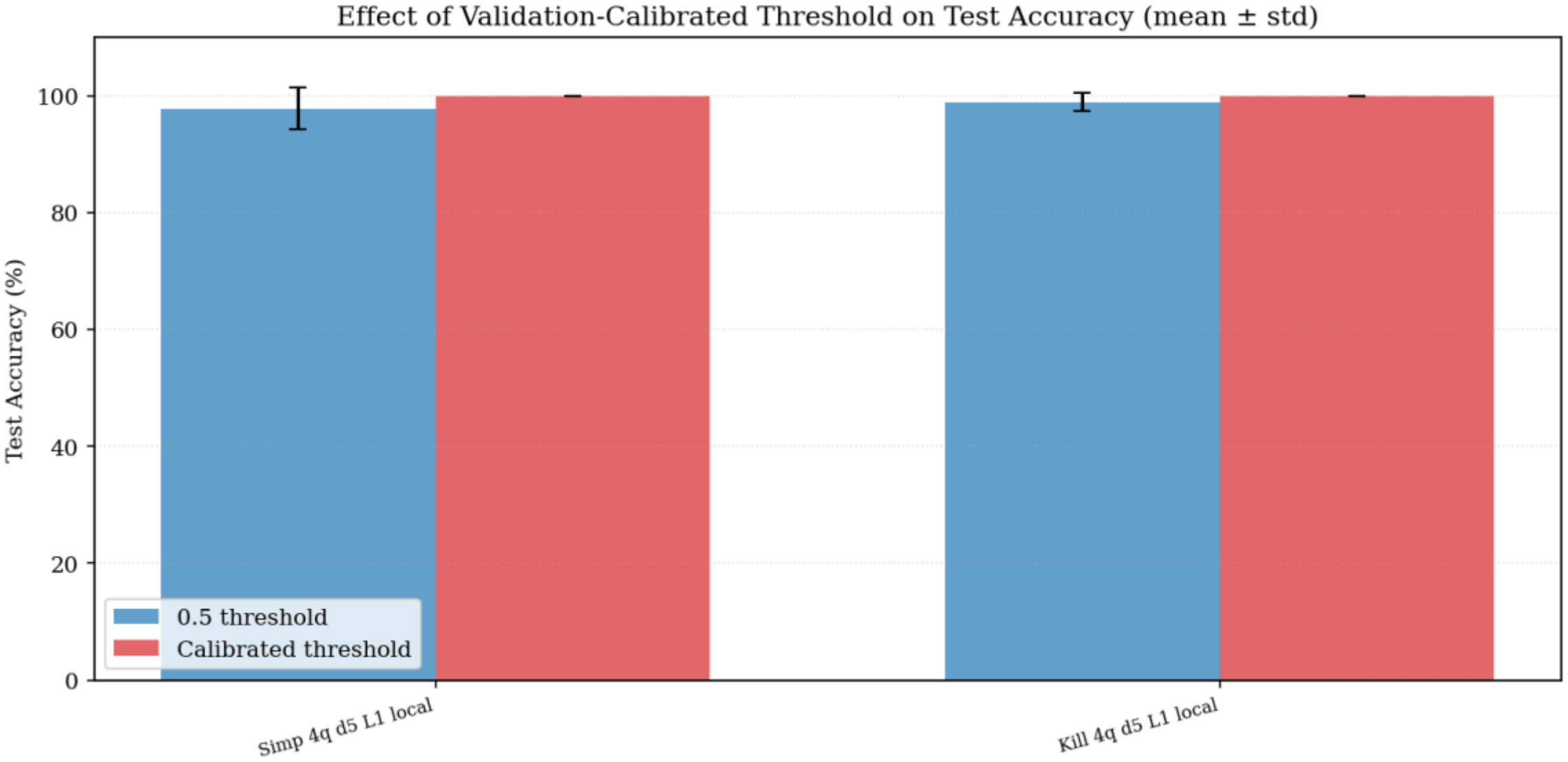

**Fig. 13** Effect of validation-based threshold calibration on test accuracy for the four-qumode models. Selecting the decision threshold on the validation set lifts the mean test accuracy from roughly 98% at the default 0.5 threshold to 100% across all seeds.

## 6. Discussion

The experiments above revise several expectations about how a continuous-variable quantum neural network should be built for this clinical screening task, and they do so in a direction that is both more nuanced and more favourable to a compact, deployable model than a single-seed reading would suggest. We discuss the parameter-efficiency picture, the role of threshold calibration, the trainability analysis that explains our architectural choices, and the limitations that bound these conclusions.

### 6.1 Parameter efficiency and architectural insights

The headline result is that a four-qumode simplified CV-QNN with just 18 trainable parameters attains the highest validation AUC of any model we tested (97.67 ± 0.06%), surpassing a 55-parameter classical network by a small but statistically consistent margin and matching every model at 100% calibrated test accuracy. The route to this result is informative. Circuit width (the number of qumodes) is the single most valuable resource: moving from two to four qumodes is worth nearly five AUC points, far more than is gained from deepening the Fock-space cutoff or stacking additional layers. Depth and a very high cutoff, by contrast, deliver diminishing returns once the model already trains. The most consequential correction to our earlier, single-seed understanding concerns the simplified versus full Killoran layer. The relationship is not a uniform victory for either design but a width-dependent crossover: the full Killoran layer's extra interferometer and squeezing gates provide a genuine, statistically significant advantage when the circuit is narrow (two qumodes), but that advantage evaporates at four qumodes, where the leaner simplified layer wins outright while using 44% fewer parameters. The practical guidance that follows is concrete: prefer the simplified layer once the circuit is wide enough to furnish an adequate feature space, and spend any additional parameter budget on width before depth. For settings with an extremely tight budget, the two-qumode simplified model remains a viable lightweight option (93.51 ± 0.77% AUC at only 8 parameters), with the understanding that at that width a small classical network is still the stronger choice.

### 6.2 Calibration and threshold selection

A recurring theme across the tables is the gap between accuracy at the default 0.5 threshold and accuracy after calibration. Because the held-out test set is small and somewhat class-imbalanced, the default threshold can leave a few borderline cases misclassified even when the underlying ranking is essentially perfect, as reflected by the high validation AUC. Selecting the operating point on the validation set and transferring it unchanged to the test set lifts the test accuracy of the four-qumode models from roughly 98% to a full 100% across all six seeds (Figure 13). This separation of ranking quality from threshold placement is why we foreground the AUC as the primary metric: it isolates the model's discriminative power from the orthogonal, and easily corrected, question of where to set the decision boundary. For a screening application this is the desirable behaviour, since the operating point can be tuned to the clinically appropriate sensitivity without retraining.

### 6.3 Mitigating the barren plateau problem in CV-QNN training

The architectural choices above (a low-dimensional PCA encoding, a restricted gate set and a single shallow layer) are not merely economical; they are what keep the model trainable at all. Variational quantum circuits are prone to barren plateaus, regions of the loss landscape where gradients vanish exponentially with system size, leaving gradient-based optimisation stranded. To quantify this we measured the variance of the loss gradients under two encodings: the full pipeline used in early experiments (a 1024-dimensional feature encoding driving a two-layer full Killoran circuit, 64 parameters) and the restricted pipeline adopted here (a 16-dimensional PCA encoding driving a single-layer simplified circuit, 18 parameters). The contrast is stark (Table 7, Figure 14). The full pipeline's gradient variance is $2.98 \times 10^{-63}$, with a mean absolute gradient of $8.74 \times 10^{-33}$, numbers at the level of machine zero, indicating a circuit whose parameters receive essentially no learning signal. The restricted pipeline raises the gradient variance to $1.81 \times 10^{-5}$ and the mean absolute gradient to $1.17 \times 10^{-3}$, an increase of roughly 58 orders of magnitude in variance. This enormous gap explains the trainability failures observed in the cutoff sweep of Section 5.2: when the effective Hilbert space is too large (whether through a high-dimensional encoding or an over-truncated, expressivity-starved one), the optimiser sits on a plateau and the training accuracy never leaves chance level. Compressing the input with PCA, restricting the gate set and keeping the circuit shallow concentrates the model on a low-dimensional, information-rich subspace where gradients remain healthy and optimisation proceeds. Parameter efficiency and trainability are thus two sides of the same design principle rather than competing objectives.

*Table 7 Loss-gradient statistics under the full and restricted encodings. Compressing the input to 16 PCA components and restricting the gate set raises the gradient variance by roughly 58 orders of magnitude, lifting the model out of the barren-plateau regime.*

| Encoding / circuit | #Params | Grad. variance | Grad. abs. mean | Grad. norm mean |
|---|---|---|---|---|
| 1024-feature, full Killoran (2 layers) | 64 | $2.98 \times 10^{-63}$ | $8.74 \times 10^{-33}$ | $5.49 \times 10^{-32}$ |
| PCA-16, simplified (1 layer) | 18 | $1.81 \times 10^{-5}$ | $1.17 \times 10^{-3}$ | $2.03 \times 10^{-2}$ |

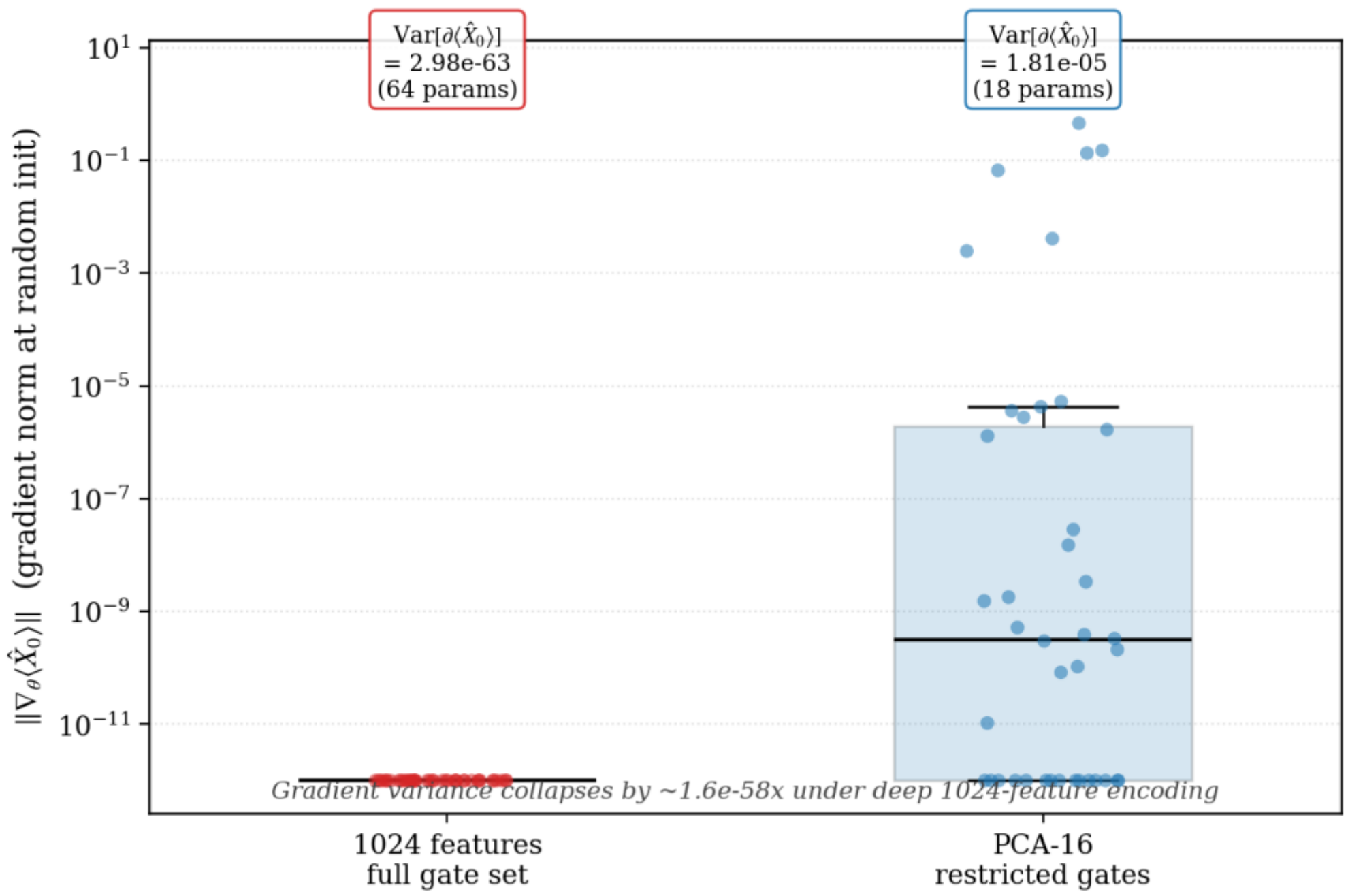

**Fig. 14** Loss-gradient magnitudes under the full 1024-feature encoding and the restricted PCA-16 encoding. The restricted pipeline escapes the barren plateau, with gradients many orders of magnitude larger than those of the full pipeline.

### 6.4 Limitations and outlook

Several limitations temper these conclusions. All quantum models are evaluated in classical Fock-space simulation rather than on photonic hardware, so the effects of finite squeezing, photon loss and detector noise on a physical device remain to be characterised. The held-out test set is small, which is why we emphasise the seed-averaged validation AUC and calibrated accuracy and report full per-seed distributions rather than relying on a single test number; the perfect calibrated test accuracies should be read as evidence of strong, consistent ranking on this cohort rather than as a guarantee of flawless generalisation. The ablation studies of Sections 5.2–5.4 are conducted at a single fixed seed to isolate each factor, so their precise numerical values carry the variance typical of a single run even though the qualitative trends are robust. Finally, the experiments use one dataset and one PCA pre-processing pipeline; confirming that the width-dependent crossover and the parameter-efficiency advantage transfer to other imaging modalities and larger cohorts is the natural next step, alongside validation on near-term photonic hardware.

## 7. Conclusion

We have presented a hybrid classical–quantum classifier for oral cancer detection from smartphone-acquired images, combining a MobileNetV1 feature extractor, principal component analysis, and a continuous-variable quantum neural network implemented on a photonic backend. The proposed simplified $\Phi \circ D \circ U_1$ architecture is most effective at four qumodes, where with only 18 trainable parameters it attains a validation AUC of 97.67 ± 0.06% (the highest of any model evaluated, classical or quantum), while reaching 100% calibrated test accuracy across all six random seeds and using 67% fewer parameters than the 55-parameter classical baseline it surpasses. For settings with an extremely tight parameter budget, a two-qumode simplified model remains a viable lightweight option (93.51 ± 0.77% validation AUC at just 8 parameters), although at that width a small classical network is the stronger choice. Beyond the headline numbers, the simplified $\Phi \circ D \circ U_1$ layer is more parameter-efficient than the full Killoran et al. (2019a) CV-QNN at every circuit width, and the comparison between them is width-dependent: the full layer retains a small but significant edge at two qumodes, while the simplified layer is significantly better at four qumodes. The restricted encoding (low-dimensional PCA inputs and a constrained gate set) raises the loss-gradient variance by roughly 58 orders of magnitude relative to a high-dimensional full-gate encoding, which is what keeps the model out of the barren-plateau regime. These results demonstrate that continuous-variable photonic quantum machine learning can achieve diagnostic performance competitive with shallow classical networks at a fraction of the parameter count, while operating in a hardware regime (room-temperature photonic processing) that is fundamentally compatible with future edge-deployable quantum inference systems. The simplified architecture and barren plateau mitigation strategies identified in this work provide concrete design guidance for building CV-QNNs that remain trainable under the parameter and energy constraints relevant to near-term photonic devices.

Several directions remain open. The present study freezes the classical feature extractor and trains only the quantum layer; extending this to a fully end-to-end optimization, in which gradients flow through both classical and quantum components, is a natural next step. All quantum experiments in this work were performed on the Strawberry Fields Fock-state simulator, which does not capture realistic photonic noise sources such as photon loss, dark counts, and finite squeezing fidelity. Characterizing the robustness of the proposed CV-QNN architecture under these noise channels, and ultimately validating the model on a physical photonic processor such as those developed by Xanadu, are essential next steps before clinical translation can be pursued. Scaling the approach to additional medical imaging tasks and larger, more diverse datasets would further strengthen the case for CV photonic quantum machine learning as a practical building block in quantum-enhanced healthcare diagnostics.

---

**Acknowledgments**

We thank the contributors of the Kaggle oral cancer dataset and the developers of the Strawberry Fields and PennyLane quantum ML libraries.

---

**References**

Arrazola JM, Bergholm V, Brádler K, et al (2021) Quantum circuits with many photons on a programmable nanophotonic chip. Nature 591:54–60. https://doi.org/10.1038/s41586-021-03202-1

Bangar S, Siopsis G, Yeter-Aydeniz K (2022) Experimentally Realizable Continuous-variable Quantum Neural Networks. In: Quantum 2.0 Conference and Exhibition. Optica Publishing Group, Boston, MA, p QTu2A.4. https://doi.org/10.1364/QUANTUM.2022.QTu2A.4

Barot S, Suthar P (2020) Oral Cancer (Lips and Tongue) images. Kaggle. https://www.kaggle.com/datasets/shivam17299/oral-cancer-lips-and-tongue-images/data

Benedetti M, Lloyd E, Sack S, Fiorentini M (2019) Parameterized quantum circuits as machine learning models. Quantum Sci Technol 4:043001. https://doi.org/10.1088/2058-9565/ab4eb5

Bharti K, Cervera-Lierta A, Kyaw TH, et al (2022) Noisy intermediate-scale quantum algorithms. Rev Mod Phys 94:015004. https://doi.org/10.1103/RevModPhys.94.015004

Braunstein SL, Van Loock P (2005) Quantum information with continuous variables. Rev Mod Phys 77:513–577. https://doi.org/10.1103/RevModPhys.77.513

Cangelosi R, Goriely A (2007) Component retention in principal component analysis with application to cDNA microarray data. Biol Direct 2:2. https://doi.org/10.1186/1745-6150-2-2

Cerezo M, Sone A, Volkoff T, et al (2021) Cost function dependent barren plateaus in shallow parametrized quantum circuits. Nat Commun 12:1791. https://doi.org/10.1038/s41467-021-21728-w

Choe S, Perkowski M (2022) Continuous Variable Quantum MNIST Classifiers—Classical-Quantum Hybrid Quantum Neural Networks. JQIS 12:37–51. https://doi.org/10.4236/jqis.2022.122005

Cong I, Choi S, Lukin MD (2019) Quantum convolutional neural networks. Nat Phys 15:1273–1278. https://doi.org/10.1038/s41567-019-0648-8

Fati SM, Senan EM, Javed Y (2022) Early Diagnosis of Oral Squamous Cell Carcinoma Based on Histopathological Images Using Deep and Hybrid Learning Approaches. Diagnostics 12:1899. https://doi.org/10.3390/diagnostics12081899

Figueroa KC, Song B, Sunny S, et al (2022) Interpretable deep learning approach for oral cancer classification using guided attention inference network. J Biomed Opt 27:015001. https://doi.org/10.1117/1.JBO.27.1.015001

Geum D-H, Roh Y-C, Yoon S-Y, et al (2013) The impact factors on 5-year survival rate in patients operated with oral cancer. J Korean Assoc Oral Maxillofac Surg 39:207. https://doi.org/10.5125/jkaoms.2013.39.5.207

Grant E, Wossnig L, Ostaszewski M, Benedetti M (2019) An initialization strategy for addressing barren plateaus in parametrized quantum circuits. Quantum 3:214. https://doi.org/10.22331/q-2019-12-09-214

Havlíček V, Córcoles AD, Temme K, et al (2019) Supervised learning with quantum-enhanced feature spaces. Nature 567:209–212. https://doi.org/10.1038/s41586-019-0980-2

Howard AG, Zhu M, Chen B, et al (2017) MobileNets: Efficient Convolutional Neural Networks for Mobile Vision Applications. arXiv:1704.04861

Hu J, Shen L, Sun G (2018) Squeeze-and-Excitation Networks. In: 2018 IEEE/CVF Conference on Computer Vision and Pattern Recognition. IEEE, Salt Lake City, UT, pp 7132–7141. https://doi.org/10.1109/CVPR.2018.00745

Jolliffe IT, Cadima J (2016) Principal component analysis: a review and recent developments. Phil Trans R Soc A 374:20150202. https://doi.org/10.1098/rsta.2015.0202

Jubair F, Al-karadsheh O, Malamos D, et al (2022) A novel lightweight deep convolutional neural network for early detection of oral cancer. Oral Diseases 28:1123–1130. https://doi.org/10.1111/odi.13825

Khanagar SB, Alkadi L, Alghilan MA, et al (2023) Application and Performance of Artificial Intelligence (AI) in Oral Cancer Diagnosis and Prediction Using Histopathological Images: A Systematic Review. Biomedicines 11:1612. https://doi.org/10.3390/biomedicines11061612

Killoran N, Bromley TR, Arrazola JM, et al (2019a) Continuous-variable quantum neural networks. Phys Rev Res 1:033063. https://doi.org/10.1103/PhysRevResearch.1.033063

Killoran N, Izaac J, Quesada N, et al (2019b) Strawberry Fields: A Software Platform for Photonic Quantum Computing. Quantum 3:129. https://doi.org/10.22331/q-2019-03-11-129

Krantz P, Kjaergaard M, Yan F, et al (2019) A quantum engineer's guide to superconducting qubits. Applied Physics Reviews 6:021318. https://doi.org/10.1063/1.5089550

Kulkarni V, Pawale S, Kharat A (2023) A classical–quantum convolutional neural network for detecting pneumonia from chest radiographs. Neural Comput & Applic 35:15503–15510. https://doi.org/10.1007/s00521-023-08566-1

LeCun Y, Bengio Y, Hinton G (2015) Deep learning. Nature 521:436–444. https://doi.org/10.1038/nature14539

Litjens G, Kooi T, Bejnordi BE, et al (2017) A survey on deep learning in medical image analysis. Medical Image Analysis 42:60–88. https://doi.org/10.1016/j.media.2017.07.005

Lloyd S, Braunstein SL (1999) Quantum Computation over Continuous Variables. Phys Rev Lett 82:1784–1787. https://doi.org/10.1103/PhysRevLett.82.1784

Lloyd S, Mohseni M, Rebentrost P (2013) Quantum algorithms for supervised and unsupervised machine learning. arXiv:1307.0411

Mani CS, Narayana L (2024) “Oral Cancer images-Chennai Dataset”, Cancer Research and Relief Trust, Chennai, India.

Mari A, Bromley TR, Izaac J, et al (2020) Transfer learning in hybrid classical-quantum neural networks. Quantum 4:340. https://doi.org/10.22331/q-2020-10-09-340

McClean JR, Boixo S, Smelyanskiy VN, et al (2018) Barren plateaus in quantum neural network training landscapes. Nat Commun 9:4812. https://doi.org/10.1038/s41467-018-07090-4

Morovati B, Lashgari R, Hajihasani M, Shabani H (2023) Reduced Deep Convolutional Activation Features (R-DeCAF) in Histopathology Images to Improve the Classification Performance for Breast Cancer Diagnosis. J Digit Imaging 36:2602–2612. https://doi.org/10.1007/s10278-023-00887-w

Preskill J (2018) Quantum Computing in the NISQ era and beyond. Quantum 2:79. https://doi.org/10.22331/q-2018-08-06-79

Schuld M, Bocharov A, Svore KM, Wiebe N (2020) Circuit-centric quantum classifiers. Phys Rev A 101:032308. https://doi.org/10.1103/PhysRevA.101.032308

Schuld M, Killoran N (2019) Quantum Machine Learning in Feature Hilbert Spaces. Phys Rev Lett 122:040504. https://doi.org/10.1103/PhysRevLett.122.040504

Schuld M, Sinayskiy I, Petruccione F (2015) An introduction to quantum machine learning. Contemporary Physics 56:172–185. https://doi.org/10.1080/00107514.2014.964942

Sonawane AB, Swamikannan LD, Tamil L (2025) Robust Classification of Oral Cancer with Limited Training Data. arXiv:2510.01547

Swamikannan LD, Sonawane AB, Patel JS, et al (2024) Oral Cancer Detection Using Mobile Vision Technology. In: 2024 IEEE EMBS International Conference on Biomedical and Health Informatics (BHI). IEEE, Houston, TX, USA, pp 1–8. https://doi.org/10.1109/BHI62660.2024.10913489

Sze V, Chen Y-H, Yang T-J, Emer JS (2017) Efficient Processing of Deep Neural Networks: A Tutorial and Survey. Proc IEEE 105:2295–2329. https://doi.org/10.1109/JPROC.2017.2761740

Ting DSW, Peng L, Varadarajan AV, et al (2019) Deep learning in ophthalmology: The technical and clinical considerations. Progress in Retinal and Eye Research 72:100759. https://doi.org/10.1016/j.preteyeres.2019.04.003

Uthoff RD, Song B, Sunny S, et al (2018) Point-of-care, smartphone-based, dual-modality, dual-view, oral cancer screening device with neural network classification for low-resource communities. PLoS ONE 13:e0207493. https://doi.org/10.1371/journal.pone.0207493

Wang S, Fontana E, Cerezo M, et al (2021) Noise-induced barren plateaus in variational quantum algorithms. Nat Commun 12:6961. https://doi.org/10.1038/s41467-021-27045-6

Warin K, Limprasert W, Suebnukarn S, et al (2021) Automatic classification and detection of oral cancer in photographic images using deep learning algorithms. Journal of Oral Pathology & Medicine 50:911–918. https://doi.org/10.1111/jop.13227

Zhang B, Zhuang Q (2025) Energy-dependent barren plateau in bosonic variational quantum circuits. Quantum Sci Technol 10:015009. https://doi.org/10.1088/2058-9565/ad80bf